\documentclass[11pt, superscriptaddress, nofootinbib, preprint]{revtex4}
\usepackage{comment}
\usepackage{slashed}
\usepackage{latexsym}
\usepackage{graphics}
\usepackage{bm}
\usepackage{subcaption}
\usepackage{wrapfig}
\usepackage[demo]{graphicx}

\usepackage{graphicx} 
\usepackage{float}
\usepackage{color}
\usepackage{amsmath}
\usepackage{amssymb}
\usepackage{feynmp-auto}
\usepackage[compat=1.1.0]{tikz-feynman}
\usepackage{slashed}
\usepackage[normalem]{ulem}

\usepackage{caption}
\usepackage{subcaption}
\usepackage{multirow}

\usepackage{color,hyperref}
\hypersetup{colorlinks,breaklinks,
            linkcolor=blue,urlcolor=blue,
            anchorcolor=blue,citecolor=blue}

\usepackage{tikz}
\usetikzlibrary{shapes.geometric,decorations.fractals,shadows,shapes.multipart,arrows,snakes,positioning,arrows}
\usetikzlibrary{decorations.pathmorphing,decorations.markings}
\usetikzlibrary{calc,intersections}

\usepackage{cleveref}

\newcommand{\be}{\begin{equation}}
\newcommand{\ee}{\end{equation}}
\newcommand{\bea}{\begin{eqnarray}}
\newcommand{\eea}{\end{eqnarray}}

\begin{document}
\title{Pathfinder for a High Statistics Search for Missing Energy  In Gamma Cascades }
\author{James B.~Dent} 
\email{jbdent@shsu.edu}
\affiliation{Department of Physics, Sam Houston State University, Huntsville, TX 77341, USA}
\author{Bhaskar Dutta}
\affiliation{Mitchell Institute for Fundamental Physics and Astronomy,
Department of Physics and Astronomy, Texas A\&M University, College Station, TX 77843, USA}
\author{Andrew Jastram}
\affiliation{Mitchell Institute for Fundamental Physics and Astronomy,
Department of Physics and Astronomy, Texas A\&M University, College Station, TX 77843, USA}

\author{Doojin Kim}
\affiliation{Mitchell Institute for Fundamental Physics and Astronomy,
Department of Physics and Astronomy, Texas A\&M University, College Station, TX 77843, USA}

\author{Andrew Kubik}
\affiliation{SNOLAB, Creighton Mine \#9, 1039 Regional Road 24, Sudbury, ON P3Y 1N2, Canada}

\author{Rupak Mahapatra}
\affiliation{Mitchell Institute for Fundamental Physics and Astronomy,
Department of Physics and Astronomy, Texas A\&M University, College Station, TX 77843, USA}
\author{Surjeet Rajendran}
\affiliation{Department of Physics \& Astronomy, The Johns Hopkins University, Baltimore, MD  21218, USA}

\author{Harikrishnan Ramani}
\affiliation{Stanford Institute for Theoretical Physics, Stanford University, Stanford, CA 94305, USA}

\author{Adrian Thompson}
\affiliation{Mitchell Institute for Fundamental Physics and Astronomy,
Department of Physics and Astronomy, Texas A\&M University, College Station, TX 77843, USA}

\author{Shubham Verma}
\affiliation{Mitchell Institute for Fundamental Physics and Astronomy,
Department of Physics and Astronomy, Texas A\&M University, College Station, TX 77843, USA}

\begin{abstract}
We investigate the feasibility of a high statistics experiment to search for invisible decay modes in nuclear gamma cascades using 200 kg of 
Cs(Tl) scintillators that are presently available at Texas A\&M. The experiment aims to search for missing energy  by robustly establishing the absence of a photon in a well identified gamma cascade. We report on the experimental demonstration of the energy resolution necessary for this search. Prior explorations of this detector concept focused on baryonically coupled physics that could be emitted in $E_2$ transitions. We point out that this  protocol can also search for particles  that are coupled to photons  by searching for the conversion of a photon produced in a gamma cascade into a hidden particle. Examples of these processes include the oscillation of a photon into a hidden photon and the conversion of a photon into an axion-like-particle either in the presence of a magnetic field or via the Primakoff process. This proof-of-concept apparatus appears to have the ability to search for hitherto unconstrained baryonically coupled scalars and pseudoscalars produced in $E_0$ and $M_0$ transitions.  If successfully implemented, this experiment serves as a pathfinder for a larger detector with greater containment that can thoroughly probe the existence of new particles with mass below 4 MeV that lie in the poorly constrained supernova ``trapping window'' that exists between 100 keV and 30 MeV. 
\end{abstract}
\maketitle

\section{Introduction}

There are considerable theoretical motivations to search for light, weakly coupled particles. These particles could constitute the dark matter or act as mediators between the standard model (SM) and the dark sector \cite{Green:2017ybv}. They also arise in several frameworks of physics beyond the standard model that address theoretical puzzles such as the strong CP problem \cite{Peccei:1977hh, Weinberg:1977ma, Wilczek:1977pj}
, the hierarchy problem~\cite{Graham:2015cka}, the cosmological constant problem \cite{ Graham:2017hfr, Graham:2019bfu}, and the quantum nature of gravity. A significant impediment in searching for such particles in a controlled laboratory setting is statistics: the more weakly coupled a particle, the harder it is to produce it. As a result, probes of such particles have largely centered around their effects on astrophysical bodies and cosmology.

This is clearly unsatisfactory. First, these probes are limited to the parts of parameter space where astrophysics and cosmology are well understood. Together, these place significant constraints on light particles that interact with electrons. However, particles that couple to baryons and are heavier than $\sim 100$ keV are not well constrained \cite{Green:2017ybv}. Such particles are too heavy to be produced in objects such as horizontal branch stars (HB stars). They can be  produced in supernovae and can be constrained if they interact more weakly with the standard model than neutrinos, leading to anomalous cooling. But, if their interactions with baryons are stronger than neutrinos, these particles do not efficiently cool the star and are thus unconstrained. Cosmological limits on such particles are also weak since the baryon abundance drops significantly as the universe cools below $\sim$ GeV energies. In concert with the heating of the standard model that occurs during the QCD phase transition, the relative abundance of baryonically interacting particles is suppressed. Second, these astrophysical and cosmological bounds are not robust against minor changes to the model \cite{DeRocco:2020xdt}. Additional interactions between the standard model and these particles can cause significant density-dependent effects. Given the enormous difference in densities in the early universe, the interior of stars, and the laboratory, it is not difficult to avoid these bounds when the underlying model is changed in minor ways. 

Given the need for laboratory methods to probe these particles, how can we achieve the required statistics? One way to accomplish this task using hot radioactive nuclear sources was discussed in \cite{Benato:2018ijc}. In this scheme, one searches for missing energy in a well identified gamma cascade. Consider, for example, the beta decay of $^{60}$Co to $^{60}$Ni. With a branching fraction of $99.9\%$ this decay populates the $4+$ excited state of $^{60}$Ni which decays to the ground state by first emitting a 1.17~MeV gamma to decay to the $2+$ state followed by a 1.33~MeV gamma to decay to the ground state. The idea of \cite{Benato:2018ijc} is to identify the initial 1.17~MeV gamma and with great efficiency tag the second 1.33~MeV gamma. If the second gamma was seen at a lower rate than the expected efficiency of the detector, it could imply the existence of a decay into new particles. The key parameters of this experiment are: the containment efficiency of the gammas (to avoid missing the second gamma),  the energy resolution of the setup (to separate out the gammas of interest) and the detector response time (enabling background rejection by focusing only on events within a short $\sim$ ns interval). Another central parameter is the nuclear source itself. While $^{60}$Co is a readily available source, for instance, the fact that the two gammas are close to each other in energy limits the ultimate reach of an experiment using this isotope. As discussed in \cite{Benato:2018ijc}, these limitations can be overcome using $^{24}$Na. However, the short half-life introduces experimental and logistical complications. In this work we explore an expanded list of nuclei including $^{46}$Sc. This nucleus has a 83 day half-life, sufficient energy separation between the two decay photons and importantly, the first photon is more energetic (1.12 MeV) than the second (889 keV). These features make this a promising isotope for this experiment.

The purpose of this paper is to further develop the experimental implementation of the theoretical concepts proposed in \cite{Benato:2018ijc} in a concrete experimental apparatus at Texas A\&M university. This apparatus uses CsI(Tl) crystals that act as a reasonably fast scintillating detector for the produced gammas. We discuss the properties of these crystals, the trigger logic and the optimal geometries necessary for maximum containment. We also report on an experimental demonstration of the energy resolution needed for this experiment. These inputs are  used to estimate the sensitivity of the apparatus. All of these are discussed in Section \ref{sec:setup}. Following this, in Section \ref{sec:transitions}, we consider a broader class of transitions than discussed in \cite{Benato:2018ijc}. This includes $M_1$ and $M_0$ transitions that are particularly useful in searching for pseudoscalars such as axions and axion-like-particles and $E_0$ transitions where the standard model produces $e^+/e^-$ pairs that can be tagged far more effectively than gammas. In Section \ref{sec:models}, we estimate the reach of this setup for a variety of models, well beyond the ones considered in \cite{Benato:2018ijc}. We also discuss a new process - in \cite{Benato:2018ijc}, the new physics that was being searched for involved particles that directly coupled to nuclei and were thus directly produced in the decay. Here, we discuss another possibility - suppose the decay yields the usual gamma cascade. But, it is possible that the gamma oscillates into a dark particle before interacting with the standard model resulting in missing energy. This phenomenon is possible for axion-like particles in the presence of a background magnetic field or through the Primakoff process. We discuss this new process as well in Section \ref{sec:models}.

\section{Experimental Setup}
\label{sec:setup}

\begin{figure}
\centering
\begin{subfigure}{.5\textwidth}
  \centering
  \includegraphics[width=.7\linewidth]{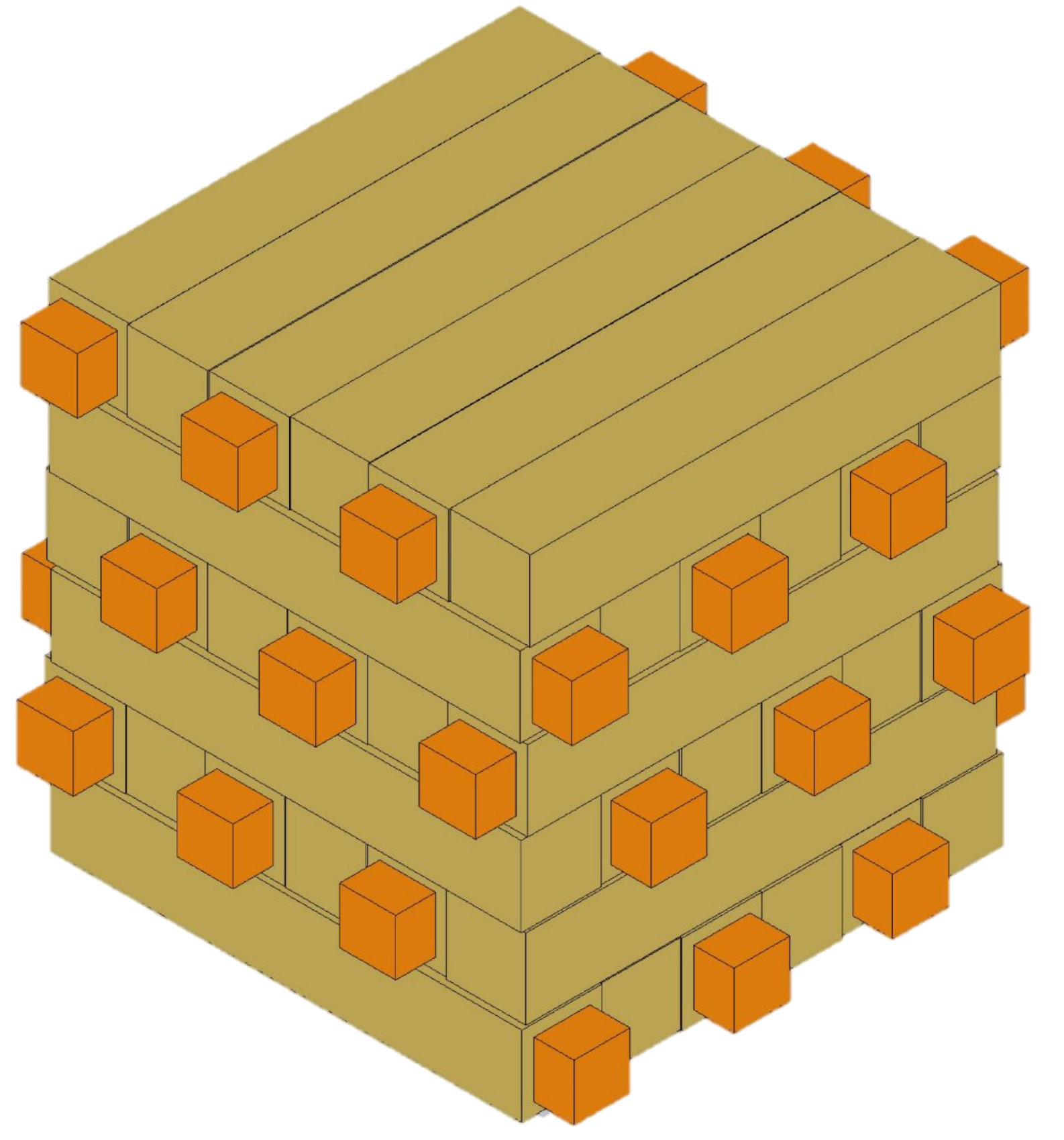}
  \caption{}
  \label{fig:sub1}
\end{subfigure}%
\begin{subfigure}{.5\textwidth}
  \centering
  \includegraphics[width=1\linewidth]{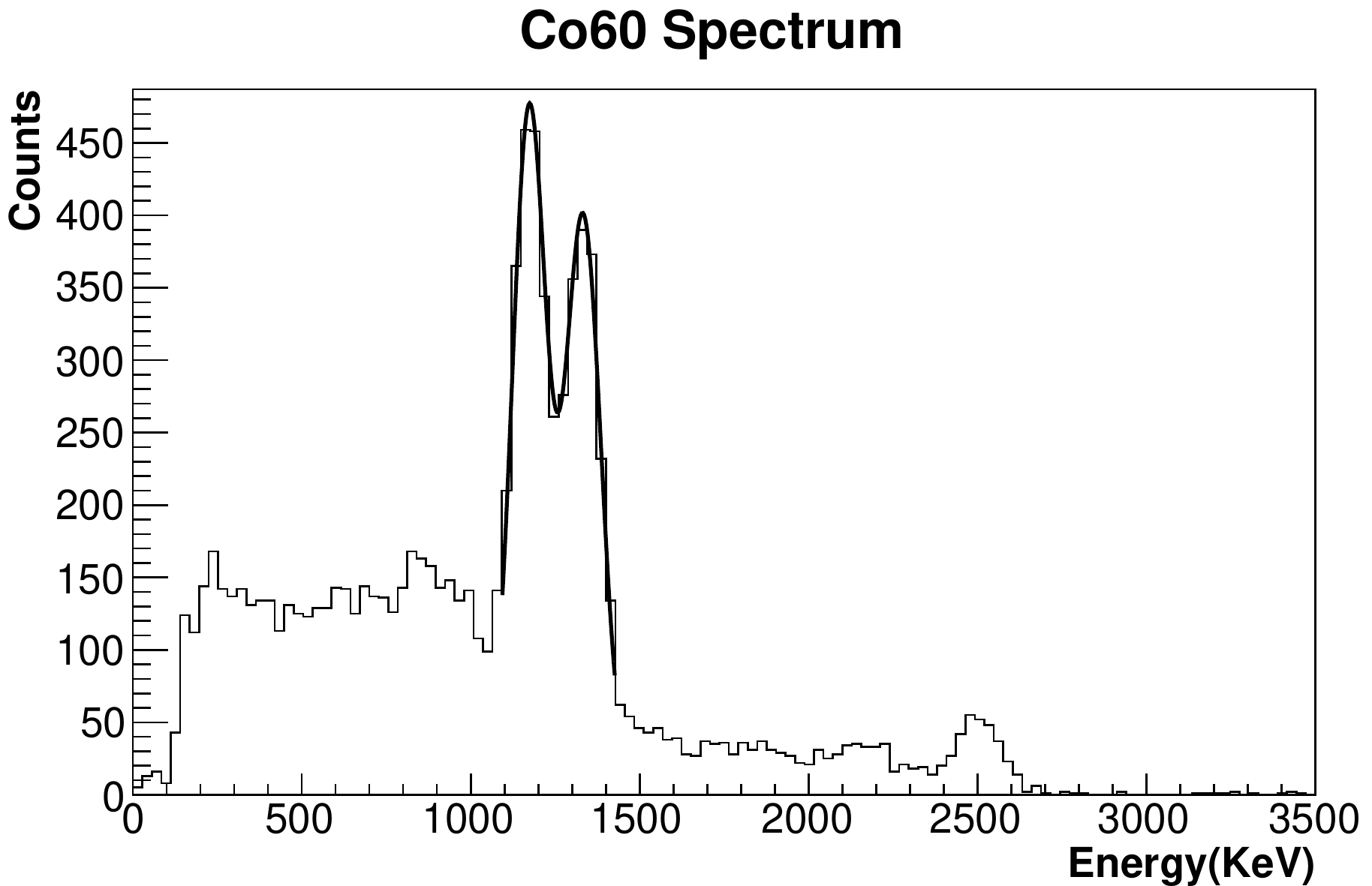}
  \caption{}
  \label{fig:sub2}
\end{subfigure}
\caption{{{\bf (a)} Conceptual drawing of a 200-kg prototype experiment with each CsI(Tl) scintillating crystal coupled to SiPM array at one end. {\bf (b)} } $^{60}$Co peak observed in a half size crystal with a PMT coupled to one end.}
\label{fig:test}
\end{figure}

The proposed experimental setup will utilize 2-5 tons of 5 kg, $ 2^{\prime\prime} \times 2^{\prime\prime} \times 12^{\prime\prime}$ in dimension CsI(Tl) crystals.  CsI(Tl) has been chosen due to its high light yield, high density and less hygroscopic nature (which makes it suitable to be used under normal atmospheric conditions).  The initial prototype will be an approximately 200 kg experiment with 36 crystals organized as a cube as shown in Fig.~\ref{fig:sub1} to provide uniform containment in a 4$\pi$ geometry. In our setup, each CsI(Tl) crystal is wrapped with a Teflon tape for improved internal reflection of the scintillating photons and then a layer of thin aluminized mylar film is wrapped which reflects external light. One end of these crystals will be coupled to conventional EMI photomultiplier tubes (PMT) for light collection. 

The preliminary characterization of the available CsI(Tl) crystals was performed using various radioactive sources primarily $^{22}$Na, $^{60}$Co and $^{137}$Cs. We found that the two gammas from $^{60}$Co (at 1.17 MeV and 1.33 MeV) could not be resolved using the $12^{\prime\prime}$ long crystals due to the length of the crystals. The crystal length  results in a smaller light collection efficiency which would increase the relative contribution of electronic noise and hence results in poor energy resolution \cite{Blucher:1986rv}. We used a half size crystal (about $6^{\prime\prime}$ in length) to have an acceptable resolution at MeV scale to resolve the two gammas from $^{60}$Co, as shown in Fig.~\ref{fig:sub2}. For future upgrades we plan to use SiPMs as photodetetcors instead of PMTs, as they have smaller profiles than the large PMTs. These facts also simplify the complexity of the geometry. The ability to search for missing gammas depends critically on how hermetic the set up is, thus any larger scale up will utilize SiPMs. Other benefits of SiPMs include reduced crosstalk, lower dark count, low afterpulse, outstanding photon detection efficiency, low voltage operation, high gain and good signal to noise ratio. SiPMs are also insensitive to magnetic fields, enabling their use in setups where magnetic fields may be necessary. The cost of SiPMs was prohibitive for this 200 kg prototype that was entirely built using internal funds. The end of the CsI(Tl) crystal that will be attached to the PMT will be well polished and examined for any possible defects and contamination. After selecting the crystals to be used for the prototype assembly, each crystal is connected to a PMT and it is then characterized for  energy response primarily using 511 keV gammas from a $^{22}$Na radioactive source.

Besides calibration, $^{22}$Na is also used for the estimation of detector efficiency for detecting gammas from the radioactive sources. $^{22}$Na decays by emitting a positron mostly into a 1275~keV level of $^{22}$Ne. These positrons annihilate with the surrounding electrons, producing two 511~keV photons. If we tag one of these two gammas and look for the other gamma in the same event,  we can perform an in-situ measurement of the gamma detection efficiency of the setup. The new physics that is being searched for in this experiment will in general couple very differently to nucleons and electrons. In particular, the electron couplings are much more constrained. Thus if we see missing gammas in the nuclear decay channel without any loss in the positron annihilation channel, it will be convincing evidence of a genuine signal. We also propose to use this setup to probe axion-like-particles by searching for the disappearance of photons via the Primakoff process or in the presence of an applied magnetic field. The verification of detector efficiency for this channel requires a different strategy. In the case where the conversion happens due to an external magnetic field, the signal can be turned off by reducing the magnetic field and this can be used to prove the existence of new physics. The Primakoff case is more challenging - the verification of the signal could potentially be performed by utilizing the fact that the conversion efficiency is independent of the photon energy whereas a loss channel in the detector is likely to be energy-dependent.

 \begin{table}[h]
    \centering
    \begin{tabular}{|c|c|c|c|c|}
     \hline
      Quantity    & CsI(Tl) & NaI(Tl) & BGO \\ \hline
      Emission spectrum max (nm) & 550 & 415 & 480 \\
      Density (g/cm$^3$) & 4.51 & 3.67 & 7.13 \\
      Hygroscopic & Slightly & Yes & No \\
      Refractive index (at max. emission) & 1.79 & 1.85 & 2.15 \\
      Radiation length (cm) & 1.85 & 2.59 & 1.12 \\
      Interaction length (g/cm$^2$) & 167 & 152 & 156 \\
      Light yield (photons/keV$\gamma$) & 54 & 38 & 10 \\
      Decay time ($\mu$s) & 0.679 (fast), 3.34 (slow) & 0.250 & 0.300 \\
      \hline
    \end{tabular}
    \caption{Properties of CsI(Tl) along with other scintillating materials \cite{Blucher:1986rv}}
    \label{tab:models}
\end{table}

We plan to put the source at the center of this 200 kg detector assembly and perform the efficiency measurements after each detector has been calibrated for gamma energy. Table \ref{tab:models} lists the properties of CsI(Tl) crystals which will be used in this experiment. The slow component of the decay time of CsI(Tl) crystals is around 3~$\mu$s. To avoid pileups, we can select the activity of the source to have an average time between decays of $\sim$~5~$\mu$s, allowing a trigger rate around 200 kHz. Pile ups can still be handled offline during data analysis as long as the two events don't occur within a ~$\mu$s. With this trigger rate, we expect to take data for a year and analyze around $10^{13}$ decays/year. This will be phase 1 of this experiment. For data acquisition, the signals from the PMTs are fed into a spectroscopic amplifier (CAEN model N568E) which also performs pulse shaping and has a web interface to control the gain and shaping time. The signal is then fed into a leading edge discriminator with multiplicity option for the trigger (CAEN model N841). VME-based 50 MHz analogue-to-digital converter with 64-channels (CAEN VX2740) is used for signal digitization and recording. This also offers multi-channel analysis for nuclear spectroscopy and it is thus well suited for our purpose. The data will be analyzed with an Optimum-filter-based analysis package and the ROOT analysis package~\cite{Brun:2003xr}. Additionally, the detector and laboratory geometry will be modeled, and backgrounds will be calibrated and validated using the GEANT4 simulation toolkit using the Shielding Physics List~\cite{Agostinelli:2002hh}. Work along this direction has already been initiated for a smaller 50~kg CsI(Tl) prototype geometry.

\section{Multipole Transitions}
\label{sec:transitions}
 The transitions considered in this work can be broadly classified into two types. In the first type, the standard model process is a beta/electron capture (EC) decay followed by a gamma cascade of two or more photons all the way to the daughter nucleus' ground state. We use all but one of the photons as the trigger and an event corresponds to exactly one photon in the cascade being absent. The transition instead proceeds through the small but non-zero branching fraction to a dark particle.
 The $E_2$ transitions considered in \cite{Benato:2018ijc} as well as the $M_1$ transition in $^{65}\textrm{Ni}$ which are well suited for axion searches are of this type. For dipole and higher transitions, the sensitivity to the rare dark decay is limited by the containment of a single photon, i.e. the smallest branching fractions that can be probed is given by, 
 \begin{equation}
 Br_{\rm dark}^{\rm lim} \approx \exp\left(-{\frac{R_{\rm det}}{\lambda_{\rm abs}}}\right). \quad \textrm{ (for dipole and higher transitions)}
 \end{equation}
 where $\lambda_{\rm abs}$ is the mean free path of the photon in the scintillator and $R_{\rm det}$ is the radius of a spherical detector. After this branching fraction is reached, subsequent statistics only improve the reach after background subtraction, and limits therefore scale as the square-root of statistics. 
 
 
 This limitation is relaxed for particles produced in  $E_0$ or $M_0$ transitions. These are rare monopole transitions which cause the decay of a $0^+$ or $0^-$ to a ground state $0^+$. Due to angular momentum conservation, this can never proceed through a single photon. In the standard model, the $E_0$ transition proceeds through internal pair production ($e^+e^-$), internal conversion (single $e^-$) or through two photons with a very small probability. $M_0$ transitions have never been observed but are expected to proceed through emission of two internal electrons or through a photon pair. The interaction of electrons with the scintillator is classical, thus the probability of missing the transition entirely is negligible. The limit on this transition is instead determined by the probability of {\it entirely} missing these rare photon pairs. 
  
  \begin{equation}
 Br_{\rm dark}^{\rm lim} \approx Br_{\rm \gamma \gamma }\exp\left(-{\frac{R_{\rm det}}{\lambda_{\rm abs}}}\right). \quad \textrm{(for monopole transitions)}
 \end{equation}

We next discuss each transition in turn. 

 \subsection{$E_2$ transitions}
 The $E_2$ transitions were treated in depth in \cite{Benato:2018ijc}. We expand on the number of candidate nuclei studied in this work. Candidate nuclei which exhibit $E_2$ transitions are tabulated in Table~\ref{tab:e2table}. Nuclei in which the second photon in the cascade has lower energy than the first trigger photon, i.e. $E_2< E_1$, are preferred because the probability of mis-tagging the second photon as the trigger photon due to Compton scattering is negligible. Branching fractions to various models of dark particles are presented in Section~\ref{sec:models} and model-dependent limit projections for $E_2$ transitions are made in Section~\ref{sec:reach}. 
 \begin{table}[]
    \centering
    \begin{tabular}{|c|c|c|c|c|}
     \hline
      Candidate  & $\tau_\frac{1}{2}$ & $E_2<E_1$? & $E_{\rm probe}$[MeV] & $E_{\rm trigger}$[MeV]   \\ \hline
      $^{207}$Bi & 31 year & Yes & 0.57 & 1.06  \\
      $^{60}$Co & 30 years & No & 1.33 & 1.17 \\
      $^{46}$Sc & 83 day & Yes & 0.89 & 1.12  \\
       $^{48}$V & 16 day & Yes & 0.98 & 1.31 \\
       $^{48}$Sc & 43.6 hr & Yes & 0.98 & 1.04 or 1.31 \\
      $^{24}$Na & 15 hr & Yes & 1.37 & 2.75 \\ \hline
    \end{tabular}
    \caption{Candidate $E_2$ transitions}
    \label{tab:e2table}
\end{table}

\subsection{$M_1$ transitions}
 $M_1$ transitions have been considered in \cite{Avignone:1988bv} in the specific context of axion searches. The decay chain of the $^{65}\textrm{Ni}$ isotope with a half-life of 2.5 hours exhibits such a transition. The parity odd-nature of these transitions makes them especially suited for searches of ALPs coupled to nuclei. Branching fractions to ALPs are presented in Section~\ref{sec:models} and model-dependent limit projections for $M_1$ transitions are made in Section~\ref{sec:reach}. 
  \begin{table}[]
    \centering
    \begin{tabular}{|c|c|c|c|c|}
     \hline
      Candidate  & $\tau_\frac{1}{2}$ & multipole & $E_{\rm probe}$[MeV]   \\ \hline
      $^{65}$Ni & 2.5 hour &  $M_1$ & 1.11 \\
      $^{90}$Nb & 14.6 hour & $E_0$ & 1.80  \\
      $^{170}$Lu & 2 day &  $M_0$ & 2.82 \\ \hline

    \end{tabular}
    \caption{Candidates for other transitions}
    \label{tab:my_label}
\end{table}

  \subsection{$E_0$ transitions}
 In the standard model, a single photon transition from $0^+ \rightarrow 0^+$ is strictly forbidden. This instead proceeds through $e^+ e^-$ emission (also called internal pair production or IPP), internal conversion followed by x-rays (IC) or through two photons (2$\gamma$).
The IPP and IC are the main decay branches. The electron loses energy classically (unlike the photon), which makes containment much more efficient. Moreover, branching fractions to invisibles such as scalars and millicharged particles are enhanced. $^{90}$Nb, whose decay has been studied in detail in \cite{pettersson1968decay,warburton1982decay}, has a 15 hour lifetime and $\beta^+$ decays predominantly to the 3.58 MeV, $8^+$ state of $^{90}$Zr. This then cascades down, and with small branching fraction populates the 1.76 MeV $0^+$ state. The subsequent $E_0$ transition to the ground state has been studied in detail in the context of $^{90}$Y $\beta^-$ decays~\cite{d2013emission}. The standard model branching fraction to two photons which forms the dominant background due to containment is estimated to be only $1.8\times 10^{-4}$ of the dominant $e^+e^-$ decay \cite{schirmer1984double}. The branching fractions to various models in $E_0$ transitions is presented in Section~\ref{sec:models} and projections in Section~\ref{sec:reach}.
 
  \subsection{$M_0$ transitions}
 Magnetic monopole transitions in decays of $0^-\rightarrow 0^+$, have been studied in a theoretical context in \cite{krutov1970higher}. Experimentally, they have been studied in light nuclei \cite{krasznahorkay2,alburger1978comment}. The heavy nucleus $^{170}$Lu is the candidate nucleus we consider for  $M_0$ transitions. In $6\%$ of its decay via electron capture, it populates the $2.82$ MeV $0^-$ state of $^{170}$Yb. The possibility of a standard model transition directly from this state to the ground state was studied in~\cite{kuhnert1993search} and no events were detected. Instead, the decay proceeds through other intermediate states with higher multipolarity. Unlike the other multi-polar transitions considered in this text, the study of $M_0$ transitions could also be novel due to the prospect of observing the $M_0$ in the standard model as well. These transitions will have favorable branching fractions to ALPs, the relevant branching fractions are discussed in Section~\ref{sec:models}  and projections for these transitions are discussed in Section~\ref{sec:reach}.
 
\section{Models and Branching Fractions}
\label{sec:models}
In this section we describe the models considered and compute the relevant branching fractions. Here we will restrict ourselves to a handful of popular models and their associated effective operators, keeping the list short, yet illustrating the breadth of models accessible to an experiment looking for missing energy in nuclear transitions.

\subsection{Dark scalars coupling to nucleons}
        We first consider a dark scalar with an effective Yukawa coupling to nucleons (specifically the proton for easy matrix element computation). 
        \begin{equation}
            \mathcal{L}\supset g_N \phi \bar{N}{N}
        \end{equation}
        In the UV-complete model, this effective interaction may come about through a Yukawa coupling to heavy or light quarks or through couplings to gluons~\cite{Knapen:2017xzo,Dev:2020eam}. 
             The effective coupling to protons can be looked for in $E$-type transitions considered in this work, i.e. $E_2$ and $E_0$ transitions. The branching fraction to a scalar, $\textrm{Br}_{\rm miss} \equiv \frac{\Gamma_\phi}{\Gamma_{\rm SM}}$ is  given by \cite{Benato:2018ijc,Izaguirre:2014cza},
        \begin{align}
        \textrm{Br}_{\rm miss}(\phi,E_2)&=\frac{g_p^2}{2e^2} \left(1-\frac{m_\phi^2}{\omega^2}\right)^{\frac{5}{2}}\nonumber \\ \textrm{Br}_{\rm miss}(\phi,E_0)&=\frac{8\pi \omega^5 }{\alpha \kappa(\omega,m_e)}  \frac{g_p^2}{e^2}\left(1-\frac{m_\phi^2}{\omega^2}\right)^{\frac{5}{2}}
        \end{align}
        
        Here
        \begin{align}
           \kappa(\omega,m)&=b\left(\frac{\omega-2m}{\omega+2m}\right)\left(\omega-2m_e\right)^3\left(\omega+2m_e\right)^2 \nonumber \\ b(S)&=\frac{3\pi}{8}\left(1-S/4-S^2/8+S^3/16-S^4/64+5S^5/512\right)
           \label{e0scalar}
        \end{align}
        The larger branching fraction in the $E_0$ case is due to favorable selection rules.

\subsection{Dark photons}
        We next consider a dark photon, $A'$, coupled to the standard model photon via kinetic mixing,  \begin{equation}
            \mathcal{L}\supset\epsilon F^{\mu\nu}F'_{\mu\nu}+ m^2_{A'} A'^2
        \end{equation}
    The branching ratio to the $A'$ is straightforward to compute for $E_2$ transitions. It is just $\epsilon^2$ with a kinematic factor that captures the effect of a non-zero $m_{A'}$ 
        \begin{align}
        \textrm{Br}_{\rm miss}(\gamma',E_2)&= \epsilon^2 \left(1-\frac{m_{A'}^2}{\omega^2}\right)^{\frac{5}{2}}\end{align}
    Note that the above expression is only valid in the limit where $m_{A'} \gg \frac{1}{\lambda_{\text{abs}}}$, the inverse mean-free-path. In this regime, the transverse mode that couples to the SM can maximally oscillate into the sterile transverse mode. In the opposite limit, i.e. for very small dark photon masses, the dark photon decouples. 
    
    For $E_0$ transitions, the transverse mode is forbidden by selection rules. The longitudinal mode behaves just like the scalar in Eqn.~\eqref{e0scalar}. However, it decouples for small $m_{A'}$ and hence the branching fraction scales as $\left(\frac{m_{A'}}{\omega}\right)^2$. Thus, we have,  
        \begin{align}
        \textrm{Br}_{\rm miss}(\gamma',E_0)&=\frac{8\pi \omega^5 }{\alpha \kappa(\omega,m_e)}  \epsilon^2 \frac{m^2_{A'}}{\omega^2}\left(1-\frac{m_{A'}^2}{\omega^2}\right)^{\frac{5}{2}}
        \end{align}
        
\subsection{Millicharged particles}
       We next consider particles carrying fractional charge under the SM photon also known as millicharge particles. This charge could arise either due to direct charge under the SM U(1), or through charge under a dark photon that kinetically mixes with the SM photon.
        The branching fractions for both cases are identical. For both $E_2$ and $E_0$ transitions, we use the calculation of matrix elements in internal pair production in the SM \cite{snover2003e+} and adapt it to milli-charge particles of mass $m_Q$ and charge $Q$ in units of the electric charge. We find
        \begin{align}
            \textrm{Br}_{\rm miss}(\textrm{milli},E_2)&= Q^2\frac{25\alpha}{9} \frac{\kappa(\omega,m_Q)}{\omega^5}\nonumber \\
             \textrm{Br}_{\rm miss}(\textrm{milli},E_0) &= Q^2 \frac{\kappa(\omega,m_Q)}{\kappa(\omega,m_e)}
        \end{align}
        
\subsection{ALP-nucleon coupling}
        Axion-like particles should generically couple to nucleons through the ``wind-coupling". The Lagrangian for the nucleon coupling can be written as,
        \begin{equation}
        \mathcal{L}\supset \frac{1}{f_{aN}} \partial_\mu a \bar{N} \gamma^\mu \gamma^5 N
        \end{equation}
        with $f_{aN}$ the relevant decay constant. Due to parity considerations, magnetic-type transitions are best suited to produce ALPs according to the following branching fraction~\cite{avignone1988search},
        \begin{align}
            \textrm{Br}_{\rm miss}(a,M_1)\sim 0.13 \left(\frac{\textrm{GeV}}{f_{aN}}\right)^2 \left(1-\frac{m_a^2}{\omega^2}\right)^{\frac{3}{2}} 
            \label{axinn}
        \end{align}
 Finally, we also consider the $M_0$ transition in the decay of $^{170}$Lu in this work. The 2.82 MeV, $0^-$ state of $^{170}$Yb is populated in $6\%$ of all the $\epsilon$ decay of $^{170}$Lu. This $0^-$ state could decay via an ALP to the $0^+$ ground state. In the SM, it instead decays via $E_2/M_1$ transitions to intermediate states. In order to make an order of magnitude estimate of the branching fraction to ALPs, we approximate the $M_0$ width to ALPs to the $E_0$ width to scalars.  
 \begin{align}
            \textrm{Br}_{\rm miss}(a,M_0)= \frac{\Gamma_{M_0}(2.82 ~\textrm{MeV})}{\Gamma_{E_2}(1.4 ~\textrm{MeV})}\approx\frac{50}{9\alpha}\frac{\rm GeV}{f_{aN}}\frac{\left((2.82 ~\textrm{MeV})^2-m_a^2\right)^\frac{5}{2}}{(1.4 ~\textrm{MeV})^5}
        \end{align}
\subsection{ALP coupling to photons}
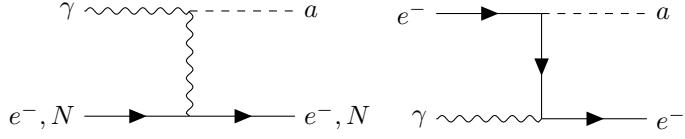
\begin{figure}
         \centering
                \begin{tikzpicture}
                      \begin{feynman}
                 \vertex (o1);
                 \vertex [right=1.4cm of o1] (f1) {\(a\)};
                 \vertex [left=1.4cm of o1] (i1){\(\gamma\)} ;
                 \vertex [below=1.4cm of o1] (o2);
                 \vertex [right=1.4cm of o2] (f2) {\(e^-,N\)};
                 \vertex [left=1.4cm of o2] (i2) {\(e^-,N\)};
        
                 \diagram* {
                   (i1) -- [boson] (o1) -- [scalar] (f1),
                   (o1) -- [boson] (o2),
                   (i2) -- [fermion] (o2),
                   (o2) -- [ fermion] (f2),
                 };
                \end{feynman}
               \end{tikzpicture}
               \begin{tikzpicture}
                      \begin{feynman}
                 \vertex (o1);
                 \vertex [right=1.4cm of o1] (f1) {\(a\)};
                 \vertex [left=1.4cm of o1] (i1){\(e^-\)} ;
                 \vertex [below=1.4cm of o1] (o2);
                 \vertex [right=1.4cm of o2] (f2) {\(e^-\)};
                 \vertex [left=1.4cm of o2] (i2) {\(\gamma\)};
        
                 \diagram* {
                   (i1) -- [fermion] (o1) -- [scalar] (f1),
                   (o1) -- [fermion] (o2),
                   (i2) -- [boson] (o2),
                   (o2) -- [ fermion] (f2),
                 };
                \end{feynman}
               \end{tikzpicture}
        \caption{Primakoff (left) and Compton-like (right) processes for photon-axion conversion.}
            \label{fig:feynman}
        \end{figure}

        ALPs coupled to photons can be produced in transitions via an internal pair production like process, with an off-shell photon producing an ALP and an on-shell photon. However, we identify a more efficient production mechanism: the decaying photon converting into an ALP in the detector. We start with the Lagrangian,
        \begin{equation}
            \mathcal{L}\supset  \frac{1}{4 f_{a\gamma}} a F_{\mu\nu}\tilde{F}^{\mu\nu} \,.
        \end{equation}
        This operator gives rise to the well-known Primakoff scattering process, in which a photon may coherently scatter with the atomic and nuclear electric fields of an atom and convert into an ALP (Fig.~\ref{fig:feynman}, left). The differential production cross-section is given by
        \begin{equation}
            \frac{d\sigma_P}{d\theta} = \frac{1}{4 f_{a\gamma}^2}\alpha Z^2F^2(t)\frac{|\vec{p}_a|^4\sin^3\theta}{t^2}\,,
        \end{equation}
        where $\theta$ is the angle of ALP with respect to the momentum direction of the incoming photon. 
        The total cross-section in the elastic limit is~\cite{Tsai:1986tx}
        \begin{equation}
            \sigma_{P} = \dfrac{\alpha }{4\pi f_{a\gamma}^2} \bigg[ Z^2 \ln (184Z^{-1/3}) + Z\ln(1194 Z^{-2/3}) + Z^2 \bigg(\ln\Big(\dfrac{403A^{-1/3} {\rm{MeV}}}{m_e}\Big) -2\bigg) \bigg]\,.
        \end{equation}
        
        Decay photons that stream into the detector volume may then undergo Primakoff conversion which gives to a probability of disappearance. This probability is given by product of an absorption probability and an effective branching fraction of photons that Primakoff-convert versus undergoing ordinary absorption; 

        \begin{equation}
        P(\gamma \to a) = \dfrac{\sigma_P}{\sigma_P + \sigma_{SM}} (1 - e^{-\sigma_{SM} n \ell}) = \dfrac{\sigma_P}{\sigma_P + \sigma_{SM}} (1 - e^{-\ell/\lambda}).
        \end{equation}
        Here, $\sigma_P$ is the total Primakoff scattering cross-section, $\sigma_{SM}$ is the total $\gamma$ absorption cross-section as a function of the photon energy $E_\gamma$, $n$ is the atomic number density and $\ell$ is the $\gamma$ path length. Alternatively, we can also express the exponent in terms of the mean-free-path $\lambda = 1/(n\sigma_{SM})$.
        
        \begin{figure}
            \centering
        \includegraphics[width=0.6\linewidth]{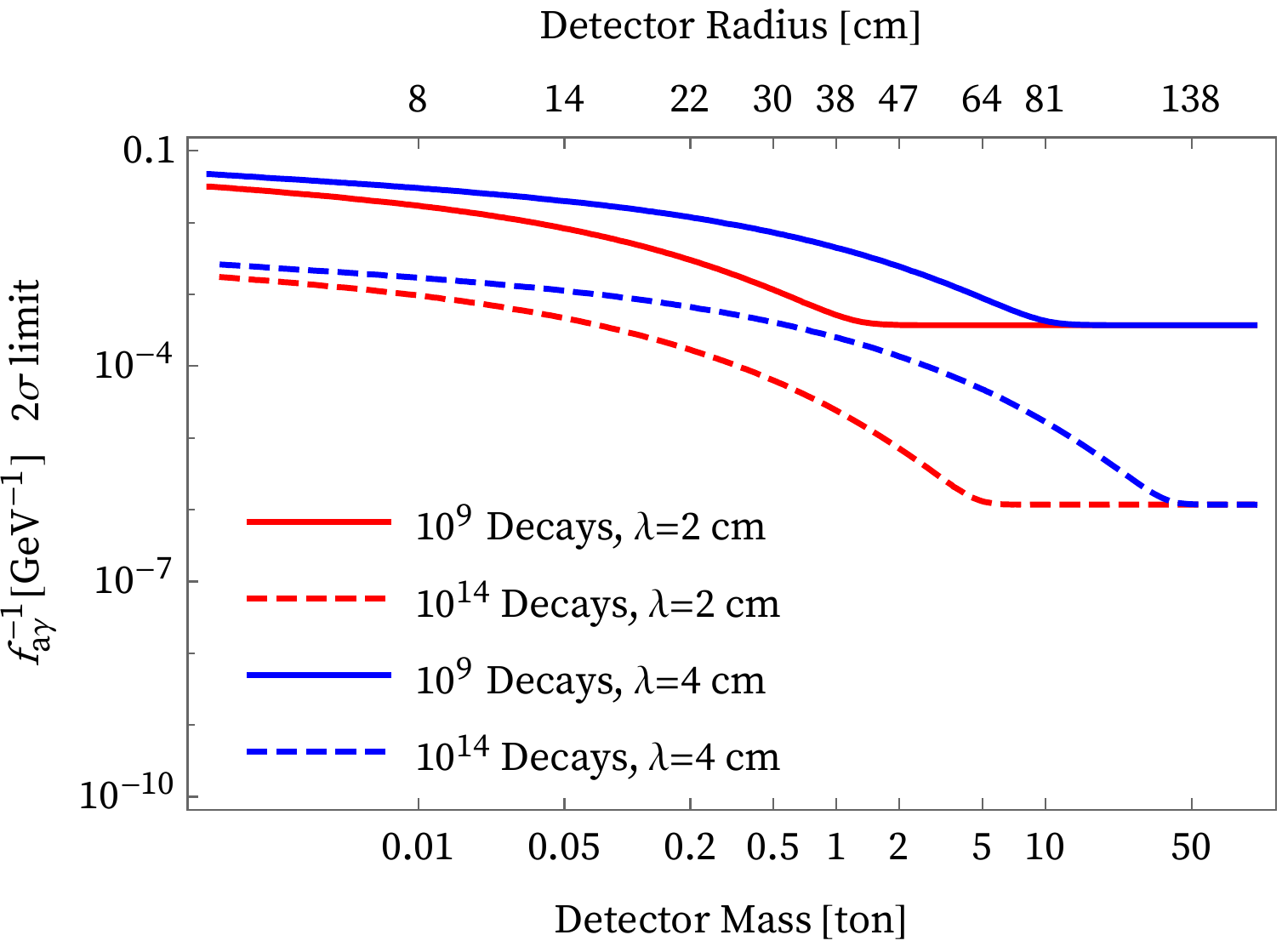}
            \caption{Sensitivity on the ALP-photon coupling, derived from Primakoff-disappearance, as a function of detector mass (bottom axis) and the equivalent detector radius (top axis) assuming spherical geometry. The effect of full containment is manifest when the sensitivity flattens out at the limit of a sufficiently large detector.}
            \label{fig:primakoff_containment}
        \end{figure}
        
        For ALPs lighter than a few keV, it is favorable to have an external magnetic field to do this conversion. There are two ways to situate this magnetic field - one could construct a geometry where the source is surrounded by a vacuum region ($\sim 1 \textrm{ m}^3$) with a large magnetic field ($B_{\rm ext}$) that is in turn hermetically covered by scintillators, or the magnetic field and the scintillator material can be co-located. The former, while more expensive, has better science reach since the photon can oscillate into an ALP over the entire volume of the vacuum region as opposed to the latter case where the conversion region is restricted to the mean free path of the photon in the scintillator. In the case of a vacuum surrounding the source, the probability for photons to convert into ALPs over a distance $L$ is given by,
       \begin{equation}
            P(\gamma\rightarrow a) = 4 \frac{ B_{\rm ext}^2 \omega^2}{m_a^4 f_{a\gamma}^2} \sin^2 \left(\frac{m_a^2 L}{4\omega}\right) .
        \end{equation}
        We will assume a magnetic field $B_{\rm ext}=10~\textrm{Tesla}$ with an extent of $L=1~\textrm{meter}$, for $m_a < 1 \textrm{eV}$,
        \begin{equation}
            P(\gamma\rightarrow a) =  \left(\frac{B_{\rm ext} L }{2f_{a\gamma}} \right)^2 \approx25 \bigg(\frac{\rm GeV}{f_{a\gamma}}\bigg)^2 \quad [{\rm Full ~Containment}].
            \label{eq:mag_primakoff}
        \end{equation}
        
        When  the scintillator material and the external magnetic field are co-located, the probability to convert to an ALP in the small mass limit is given by,
        
        \begin{equation}
            P(\gamma\rightarrow a)=4 \frac{ B_{\rm ext}^2 \omega^2}{m_a^4 f_{a\gamma}^2}\int dl  e^{-\frac{l}{\lambda}}\frac{m_a^4 l}{8\omega^2}\approx \left(\frac{B_{\rm ext} \lambda}{\sqrt{2} f_{a\gamma}} \right)^2 \quad [{\rm Prototype}].
        \end{equation}

Finally, we summarize the model operators discussed above and the relevant photon-disappearance branching ratios ($BR_{miss}$) in Table~\ref{tab:models}. In the next section, we will discuss the search potential and projected reach for each case.

\begin{table}[h]
    \centering
    \begin{tabular}{|c|c|c|c|}
     \hline
      Model  &  $\mathcal{L}_\text{int}$ &  Transition & $\textrm{BR}_{\rm miss}$    \\ \hline
      \multirow{2}{*}{Scalar (nucleon coupling)} & \multirow{2}{*}{$g_p \phi \bar{N}{N}$} & $E_2$ & $\frac{g_p^2}{2e^2} \left(1-\frac{m_\phi^2}{\omega^2}\right)^{\frac{5}{2}}$   \\
      & & $E_0$ & $\frac{8\pi \omega^5 }{\alpha \kappa(\omega,m_e)}  \frac{g_p^2}{e^2}\left(1-\frac{m_\phi^2}{\omega^2}\right)^{\frac{5}{2}}$ \\
      \multirow{2}{*}{Dark Photon} &  \multirow{2}{*}{$\epsilon F^{\mu\nu}F'_{\mu\nu}$} & $E_2$ &  $\epsilon^2 \left(1-\frac{m_\phi^2}{\omega^2}\right)^{\frac{5}{2}}$ \\
       &   & $E_0$ & $\frac{8\pi \omega^5 }{\alpha \kappa(\omega,m_e)}  \epsilon^2 \frac{m^2_{A'}}{\omega^2}(1-\frac{m_{A'}^2}{\omega^2})^{\frac{5}{2}}$  \\
      \multirow{2}{*}{Milli-charged Particle} & \multirow{2}{*}{$-Q\bar{\chi} \gamma^\mu A_\mu \chi$} & $E_2$ & $Q^2\frac{25\alpha}{9} \frac{\kappa(\omega,m_Q)}{\omega^5}$   \\
       & & $E_0$  & $Q^2 \frac{\kappa(\omega,m_Q)}{\kappa(\omega,m_e)}$    \\
      \multirow{2}{*}{ALP (nucleon coupling)}  & \multirow{2}{*}{$f_{aN}^{-1} \partial_\mu a \bar{N} \gamma^\mu \gamma^5 N$} & $M_1$  & $0.13 \left(\frac{\textrm{GeV}}{f_{aN}}\right)^2 \left(1-\frac{m_a^2}{\omega^2}\right)^{\frac{3}{2}}$  \\
       &  & $M_0$  & $\frac{50}{9\alpha}\frac{\rm GeV}{f_{aN}}\frac{\left(\omega^2-m_a^2\right)^\frac{5}{2}}{\omega^5}$  \\
      ALP (photon coupling) & $\frac{1}{4 f_{a\gamma}}  a F_{\mu\nu} \tilde{F}^{\mu\nu}$ & $E_2/M_1$ &  $\dfrac{\sigma_P}{\sigma_P + \sigma_{SM}} (1 - e^{-\ell/\lambda})$  \\
    \hline

    \end{tabular}
    \caption{Summary of the operators and branching ratios for each model considered.}
    \label{tab:my_label}
\end{table}

\section{Projected Reach}
\label{sec:reach}
In this section we estimate the potential reach of the proof-of-concept apparatus at Texas A\&M for the models considered in the previous section. We also estimate the sensitivity of a full scale experiment. All of these models are subject to constraints from existing laboratory results and astrophysical/cosmological considerations. The astrophysical limits, primarily arising from the cooling of HB stars and SN1987a \cite{Raffelt:1996wa} are highly model-dependent. For example, these limits are completely evaded in models where the particles either have a mass or coupling that depends upon the environment (see, for example, Refs.~\cite{Jaeckel:2006xm,Khoury:2003aq,Masso:2005ym,Masso:2006gc,Dupays:2006dp, Mohapatra:2006pv,Brax:2007ak,DeRocco:2020xdt}). Similar caveats also apply to cosmology, i.e. the  $N_{\rm eff}$ constraints from BBN and CMB \cite{Berlin:2020pey, DeRocco:2020xdt}. While this experiment can search for particles that are not subject to astrophysical and cosmological constraints, this model dependence encourages the production of robust laboratory limits on particles that are naively constrained by astrophysics and cosmology. 

\begin{figure}[h]
    \centering
    \begin{subfigure}[t]{0.48\textwidth}
        \centering
        \includegraphics[width=\linewidth]{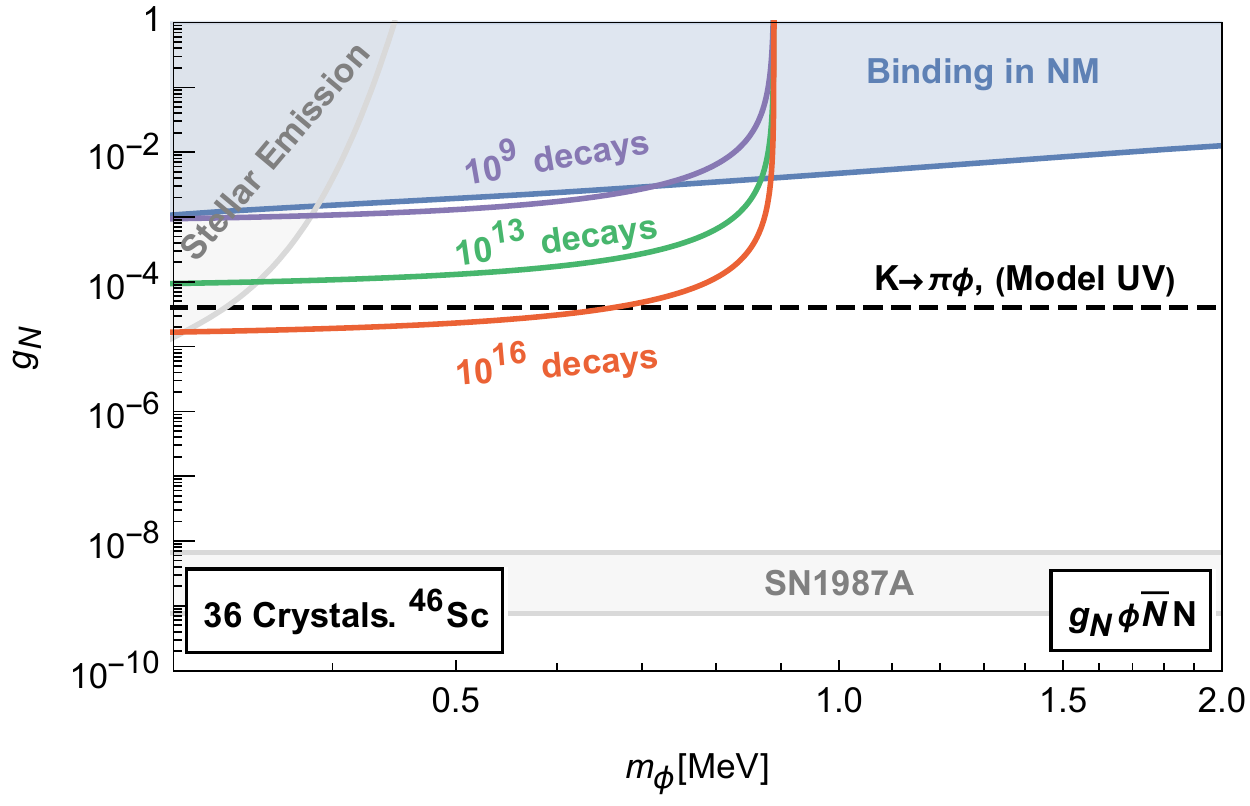} 
    \end{subfigure}
    \hfill
    \begin{subfigure}[t]{0.48\textwidth}
        \centering
        \includegraphics[width=\linewidth]{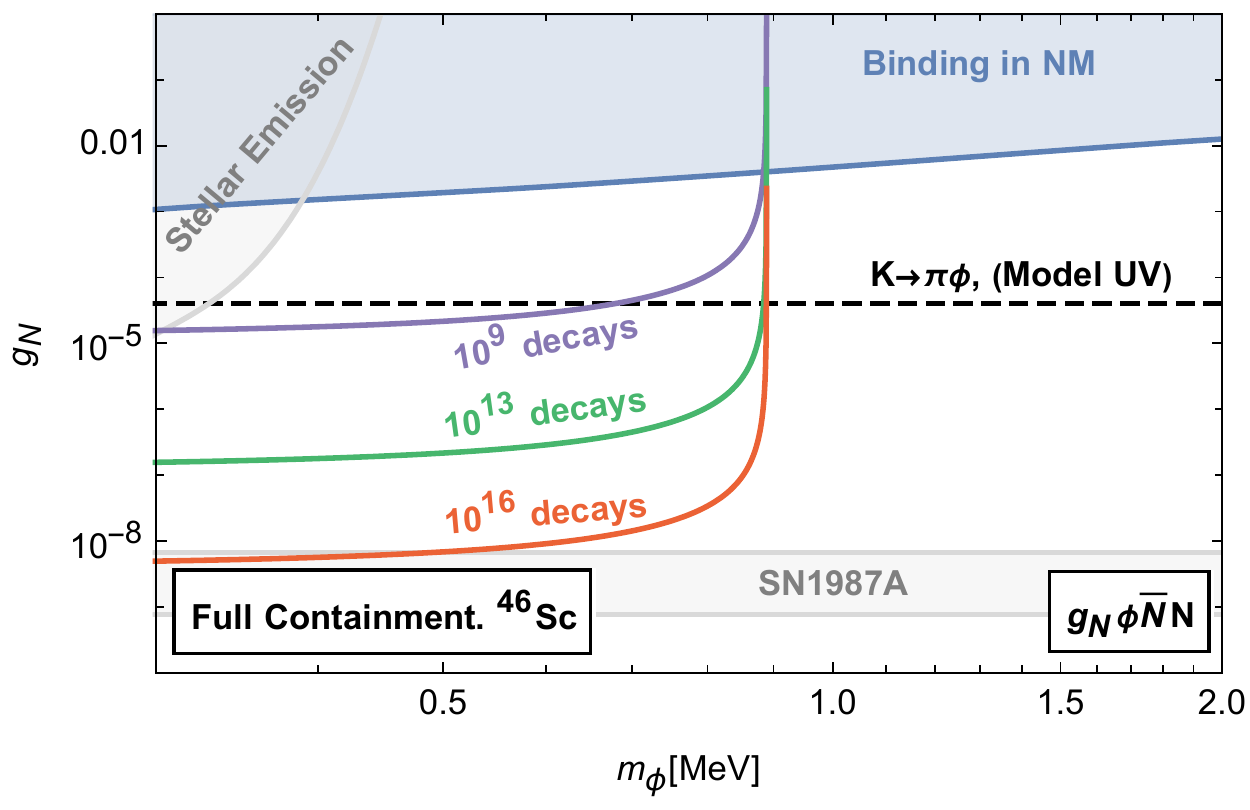} 
    \end{subfigure}
    \begin{subfigure}[t]{0.48\textwidth}
        \centering
        \includegraphics[width=\linewidth]{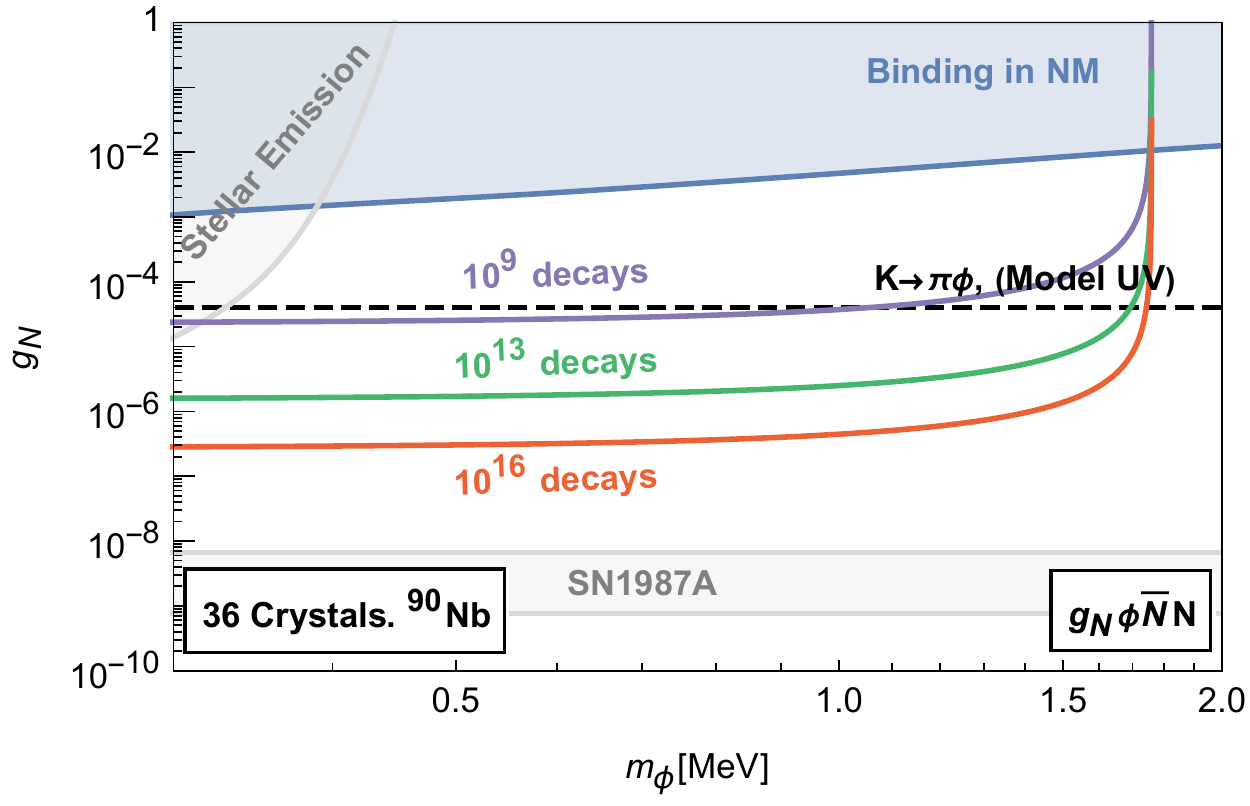} 
    \end{subfigure}
    \hfill
    \begin{subfigure}[t]{0.48\textwidth}
        \centering
        \includegraphics[width=\linewidth]{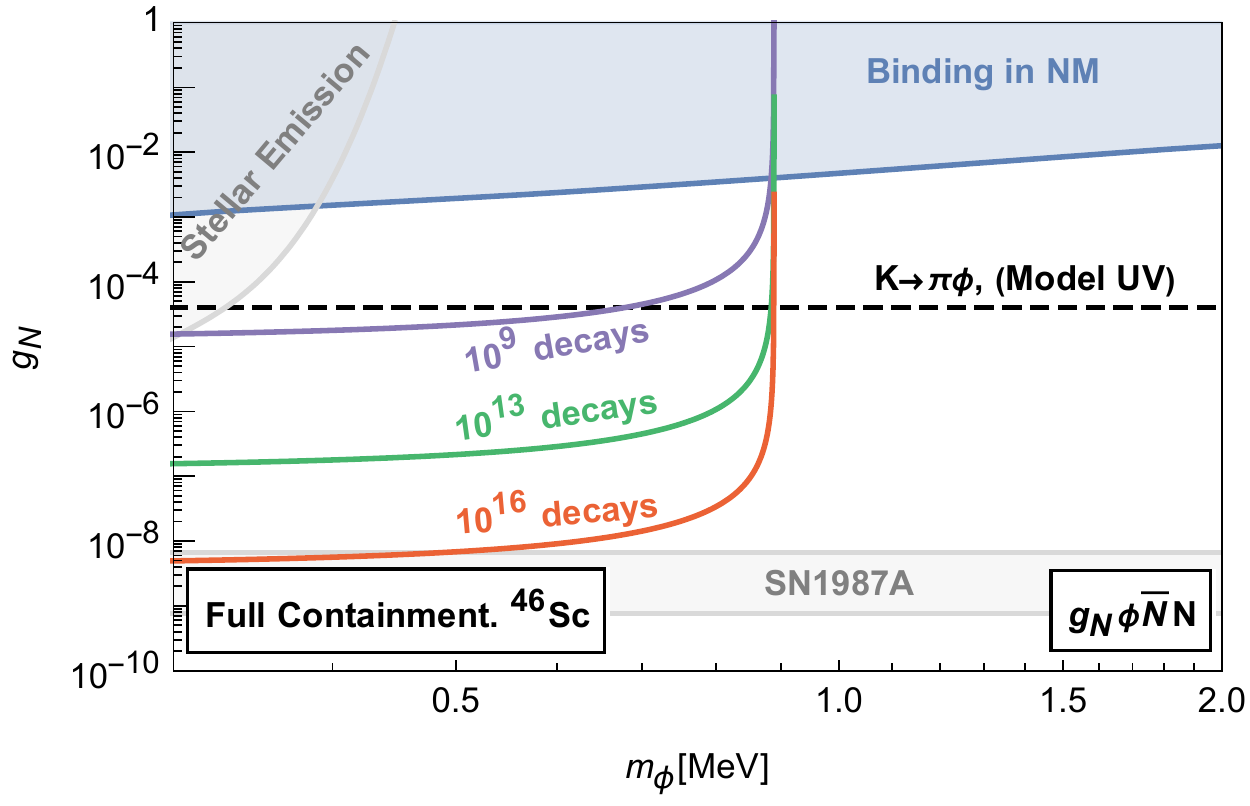} 
    \end{subfigure}
     \caption{Reach for scalar coupled to nuclei in  $E_2$ (\textbf{top}) and $E_0$ (\textbf{bottom}) transitions for current (\textbf{left panel}) and future (\textbf{right panel}) scintillator configurations.}
     \label{fig:scalar_limits}
\end{figure}

In Fig.~\ref{fig:scalar_limits} we show the projected reach over the parameter space for the scalar coupling to nucleons. The top row illustrates projections from using $^{46}$Sc decays, a candidate for $E_2$ transitions, and the bottom row illustrates projections for $^{90}$Nb, a candidate for $E_0$ transitions. Whereas the left column shows projections for 36 crystals, the right column exhibits projections if full containment of the signal photon is achieved. For large couplings in the range $g_N \sim 10^{-3} - 10^{-1}$ and for $m_\phi > 0.4$~MeV, the main constraint comes from binding energies in nuclear matter (NM)~\cite{Liu:2016qwd}. We also show astrophysical constraints from supernovae~\cite{Dev:2020eam} in addition to those from HB stars~\cite{Raffelt:1996wa}. In dotted lines we also show the limits on invisible kaon decays from E949 \cite{Artamonov:2008qb} arising from model-dependent couplings to the light quarks. 

We see that the projected reach for an exposure to $10^9$ total decays, either in the case of $^{46}$Sc or $^{90}$Nb sources, can already probe beyond the existing NM limits. Comparing the top and bottom rows, we see that there is superior reach for the $E_0$ transition in $^{90}$Nb (bottom row) compared to the $E_2$ transition. This is entirely due to the projections being containment limited in the $E_2$ transition in contrast to containment issues arising only for the severely suppressed 2-photon decay in $E_0$ transitions. This is the reason why $g_N$ scales as $N_d^{-\frac{1}{4}}$ (where $N_d$ is the total number of decays) in the prototype setup for the $E_2$ transition and scales as $N_d^{-\frac{1}{2}}$ for the other setups where containment is not an issue. We also see this by comparing the left plots to those on the right i.e. the effect of going from a 36-crystal prototype to an apparatus that offers full containment; this exponentially reduces the background of escaped gammas, thereby pushing the sensitivity curve. With $10^{16}$ decays and full containment, the entire supernova trapping window is accessible. 

\begin{figure}[h]
    \centering
    \begin{subfigure}[t]{0.48\textwidth}
        \centering
        \includegraphics[width=\linewidth]{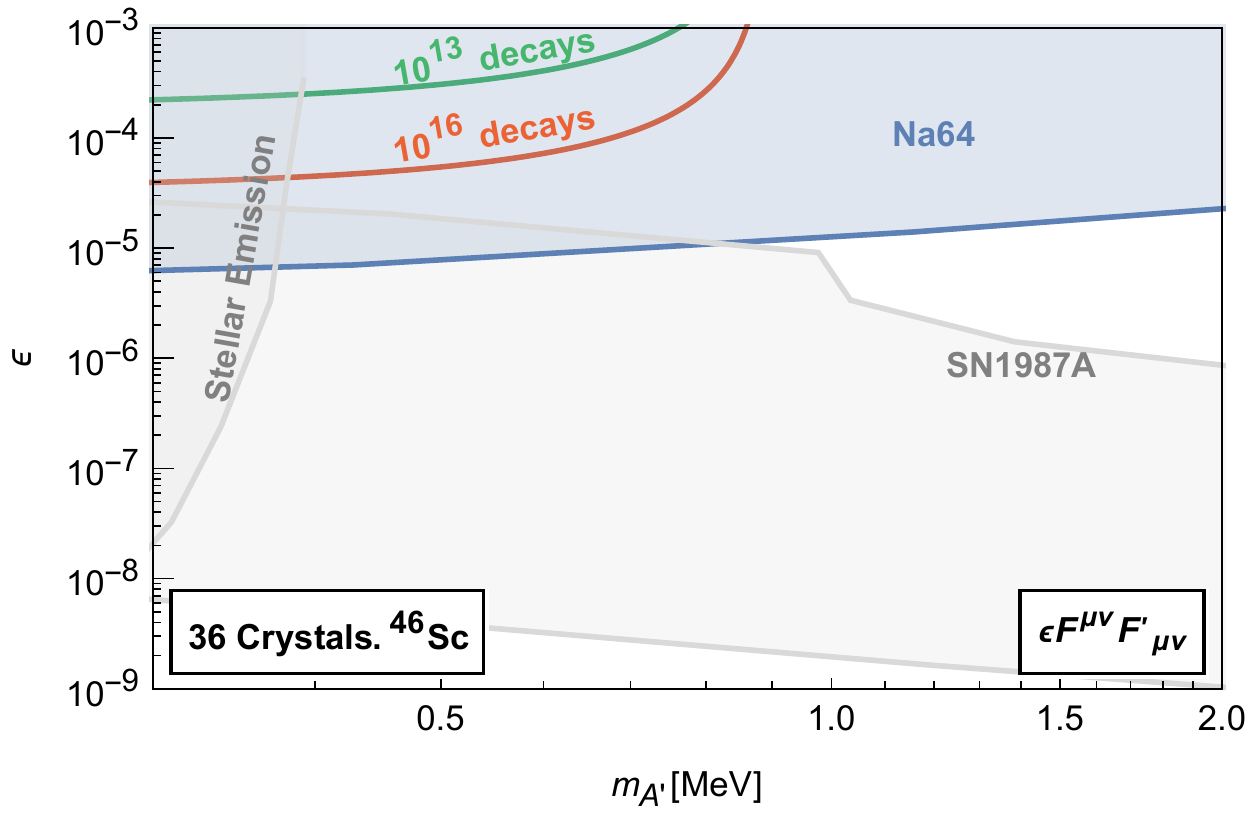} 
    \end{subfigure}
    \hfill
    \begin{subfigure}[t]{0.48\textwidth}
        \centering
        \includegraphics[width=\linewidth]{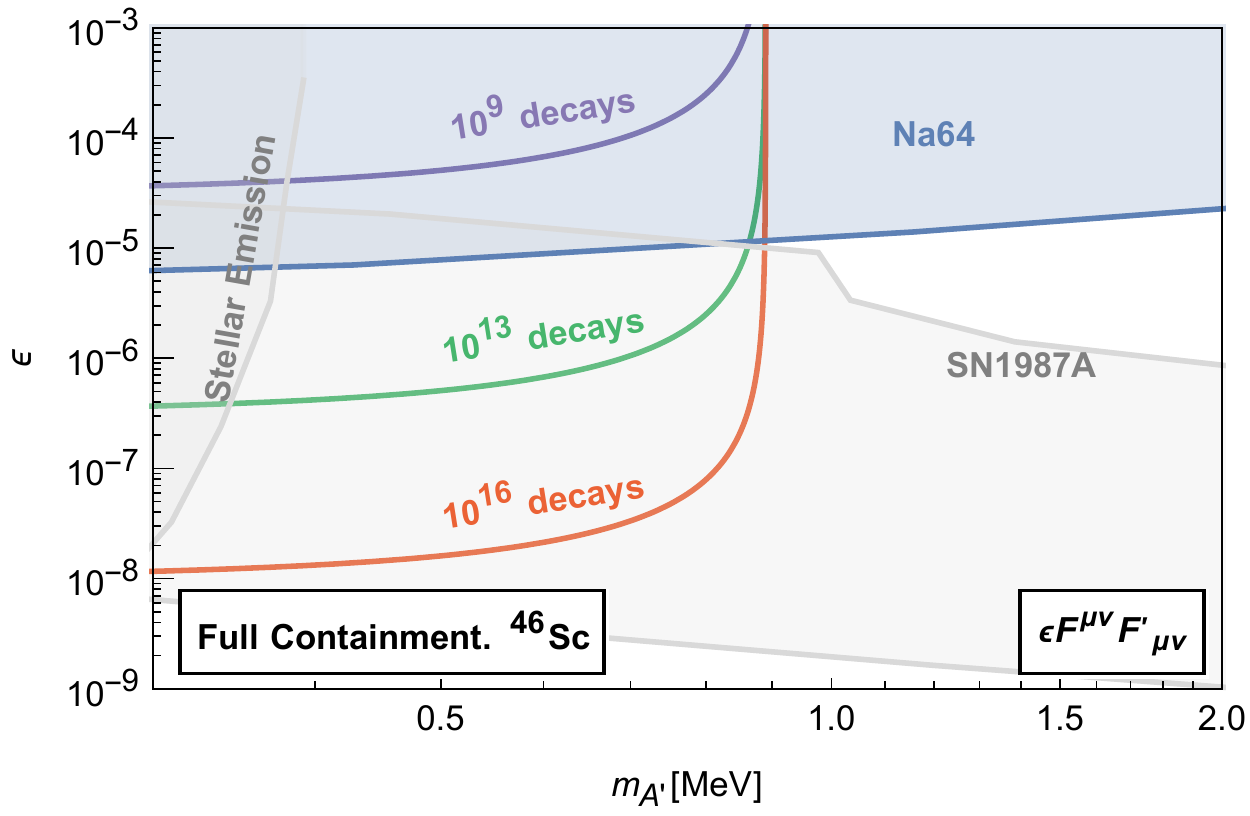} 
    \end{subfigure}
 
    \begin{subfigure}[t]{0.48\textwidth}
        \centering
        \includegraphics[width=\linewidth]{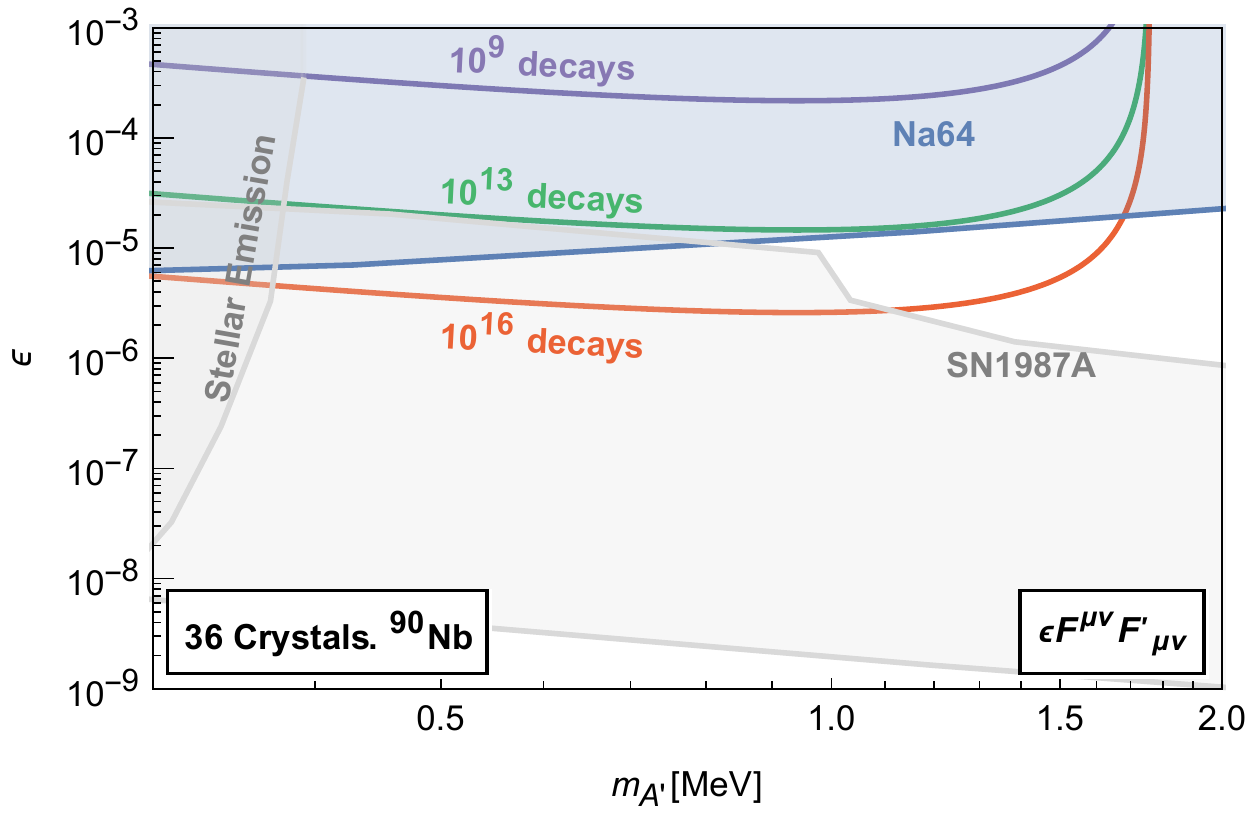} 
    \end{subfigure}
    \hfill
    \begin{subfigure}[t]{0.48\textwidth}
        \centering
        \includegraphics[width=\linewidth]{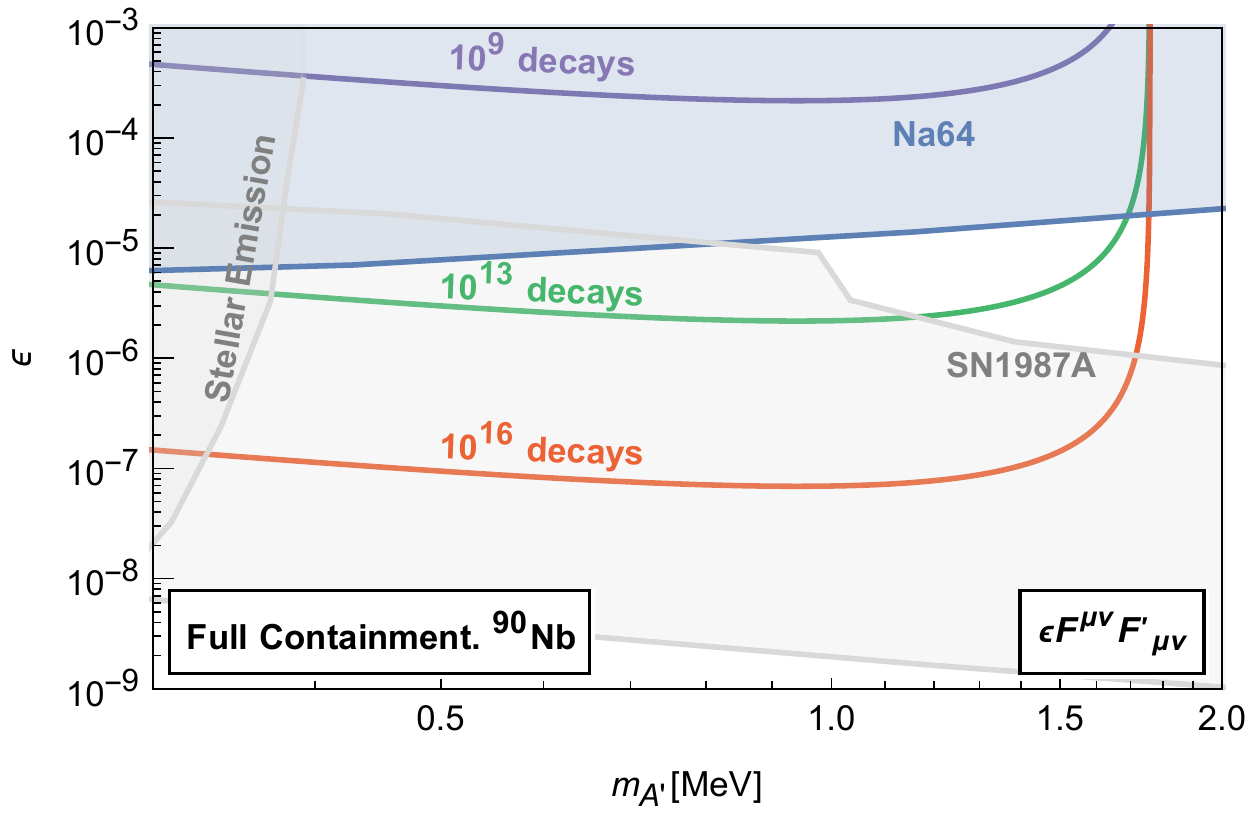} 
    \end{subfigure}
     \caption{Reach for dark photons in $E_2$ (\textbf{top}) and $E_0$ (\textbf{bottom}) transitions for current (\textbf{left panel}) and future (\textbf{right panel}) scintillator configurations}
     \label{fig:dark_photon_limits}
\end{figure}

In Fig.~\ref{fig:dark_photon_limits} we show the limits on the kinetic mixing parameter, $\epsilon$, for the dark photon model with mass $m_{A^\prime}$. The dominant laboratory bounds come from the NA64 experiment \cite{NA64:2019imj}. We show the dominant astrophysical limits~\cite{Chang:2016ntp,Hardy:2016kme} and have not plotted the sub-dominant BBN/CMB constraints (since we regard both of these as model dependent). 
The sensitivity reach here, shown again for $^{46}$Sc and $^{90}$Nb sources, suggests a strong science case for a full scale experiment with $10^{16}$ decays of $^{90}$Nb. 

\begin{figure}[h]
    \centering
    \begin{subfigure}[t]{0.48\textwidth}
        \centering
        \includegraphics[width=\linewidth]{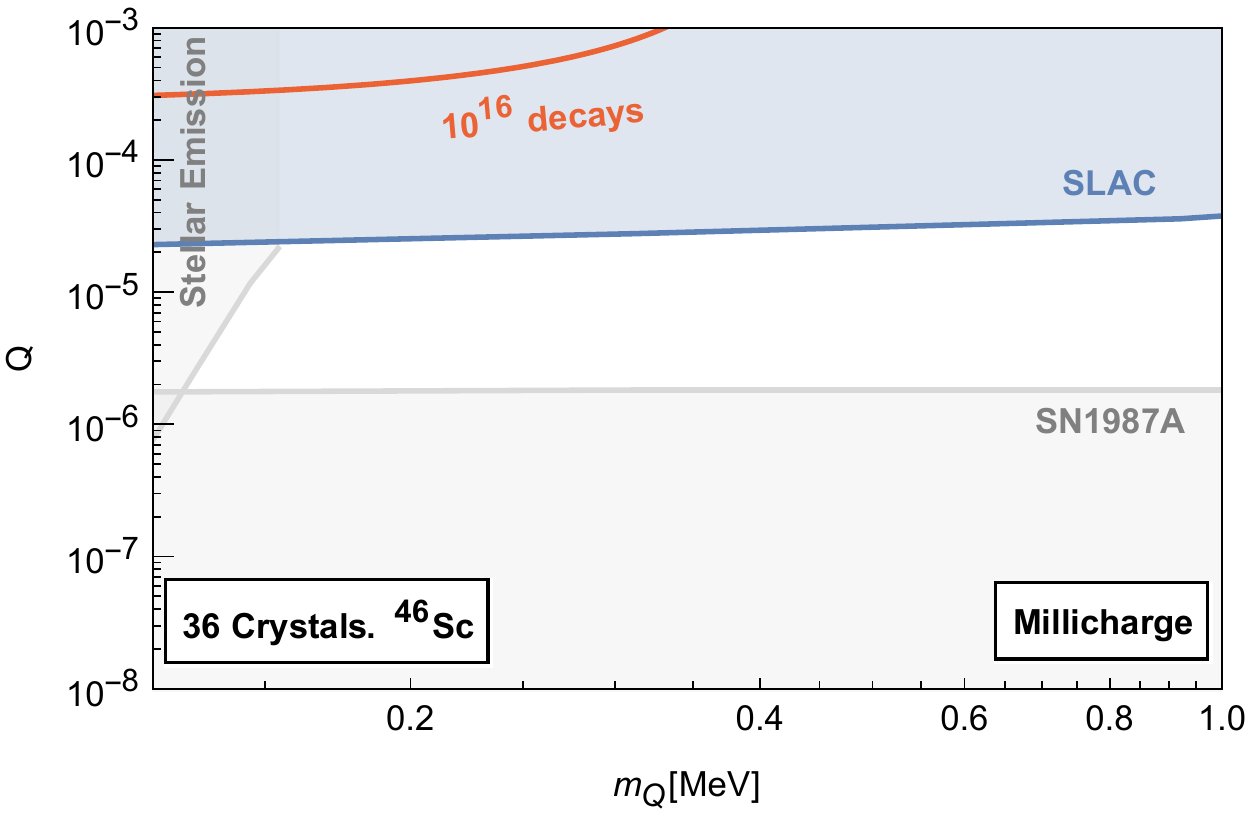} 
    \end{subfigure}
    \hfill
    \begin{subfigure}[t]{0.48\textwidth}
        \centering
        \includegraphics[width=\linewidth]{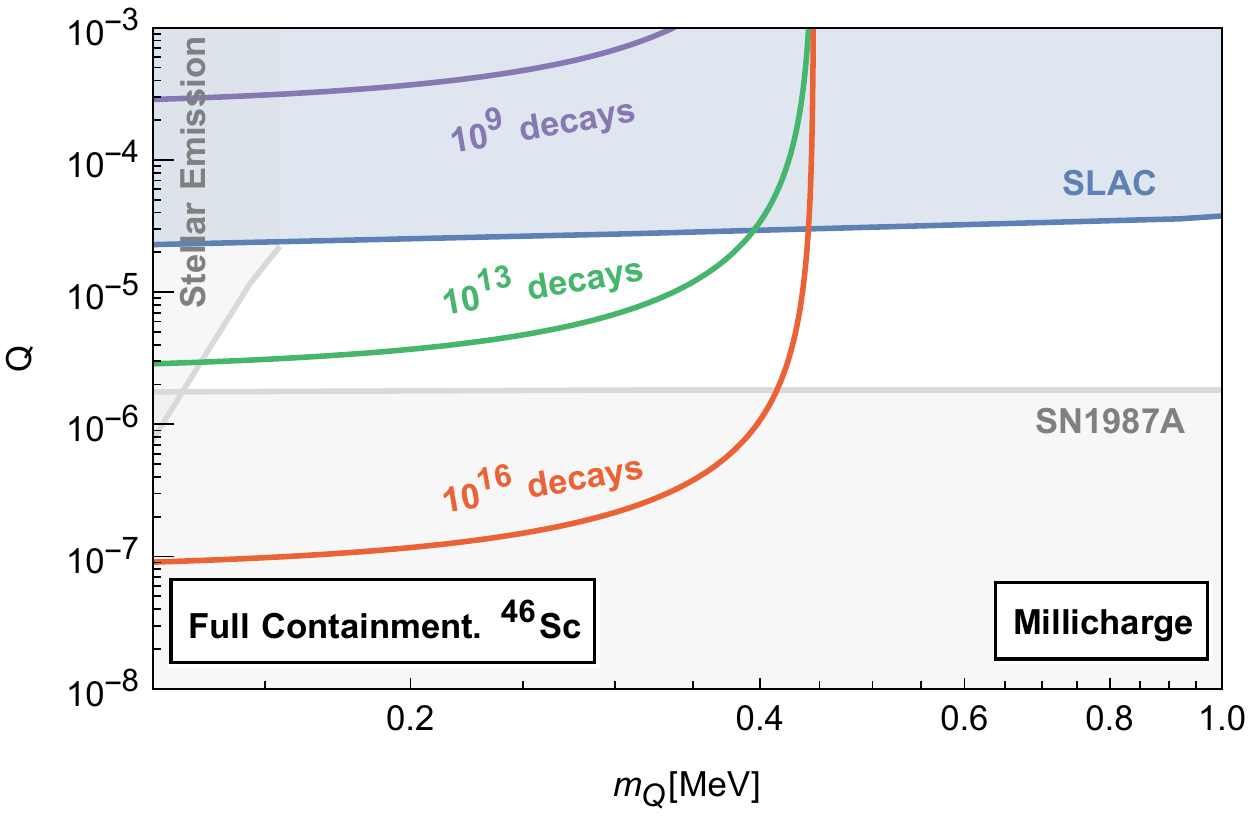} 
    \end{subfigure}
   
    \begin{subfigure}[t]{0.48\textwidth}
        \centering
        \includegraphics[width=\linewidth]{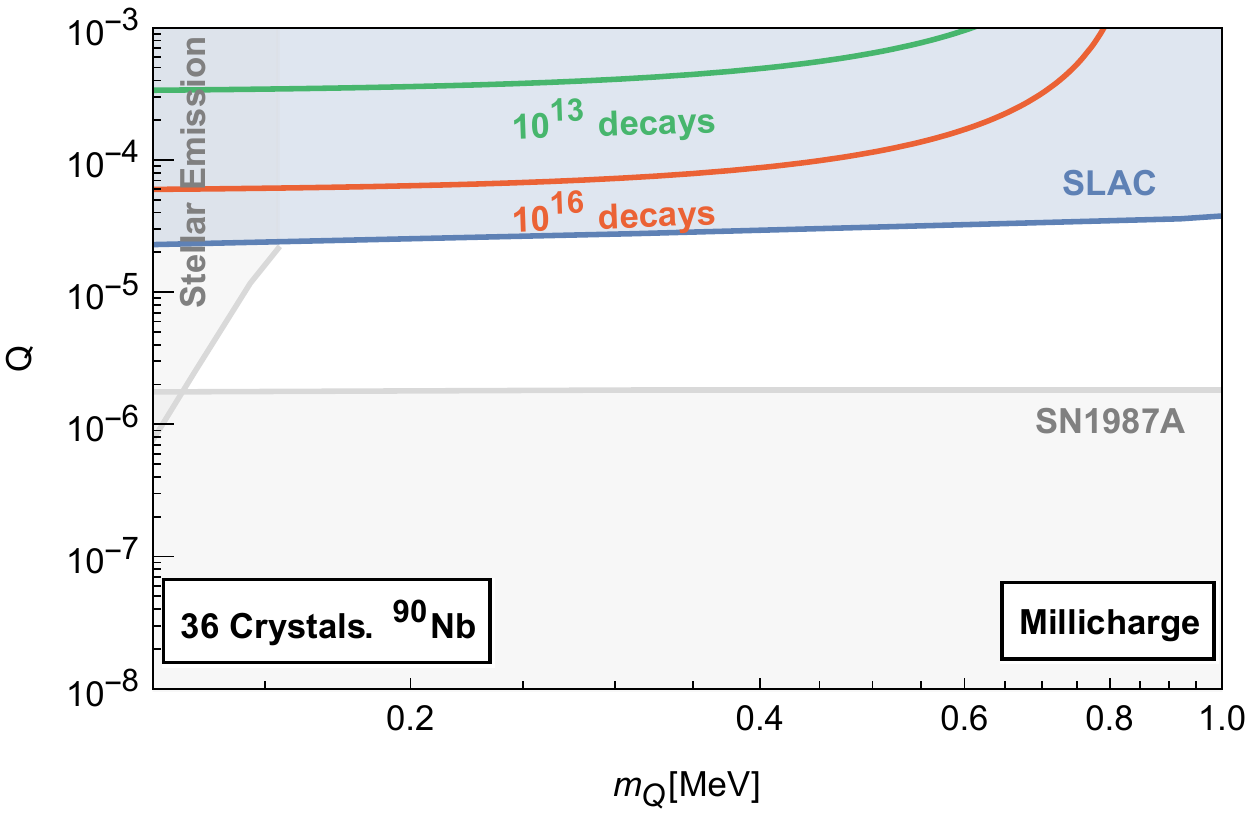} 
    \end{subfigure}
    \hfill
    \begin{subfigure}[t]{0.48\textwidth}
        \centering
        \includegraphics[width=\linewidth]{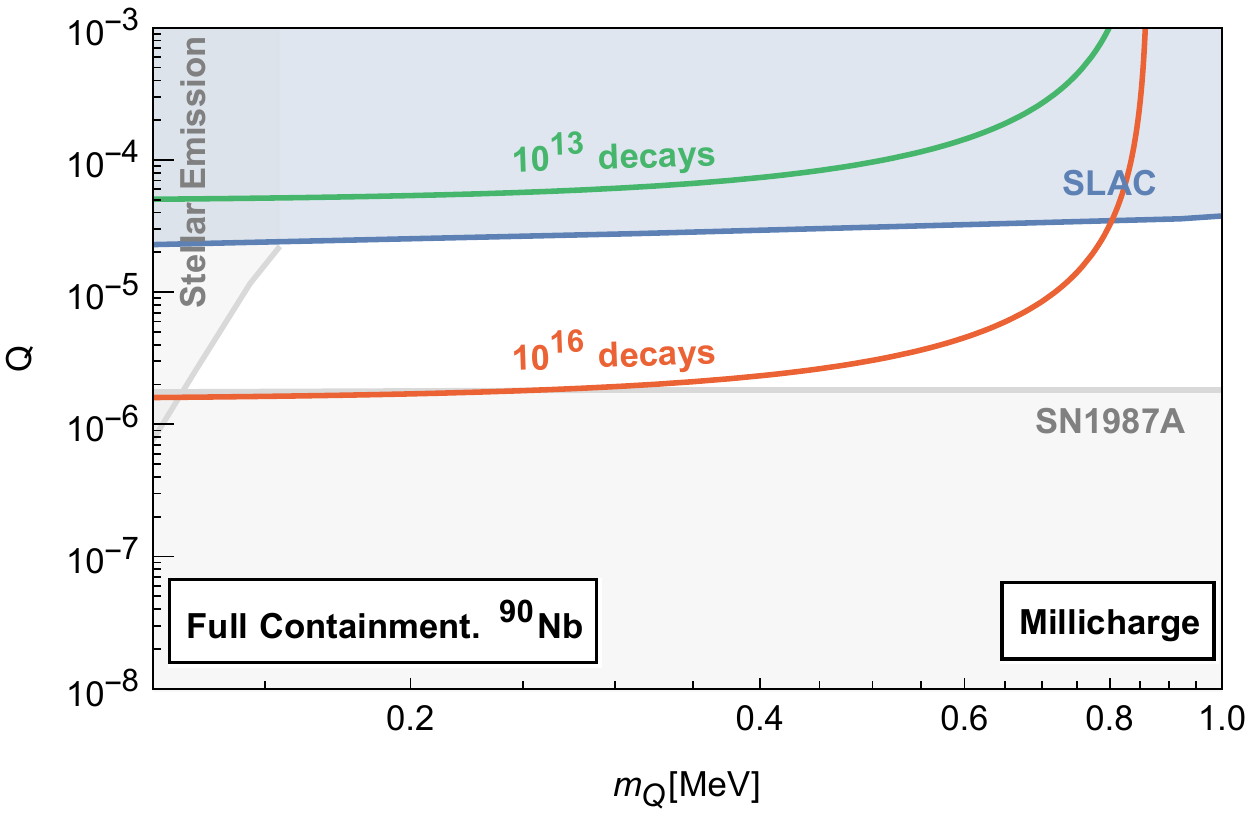} 
    \end{subfigure}
     \caption{Reach for millicharge particles in $E_2$ ($E_0$) \textbf{top} (\textbf{bottom})  transitions for current (\textbf{left panel}) and future (\textbf{right panel}) scintillator configurations}
     \label{fig:millicharge_limits}
\end{figure}

\begin{figure}[h]
    \centering
    \begin{subfigure}[t]{0.48\textwidth}
        \centering
        \includegraphics[width=\linewidth]{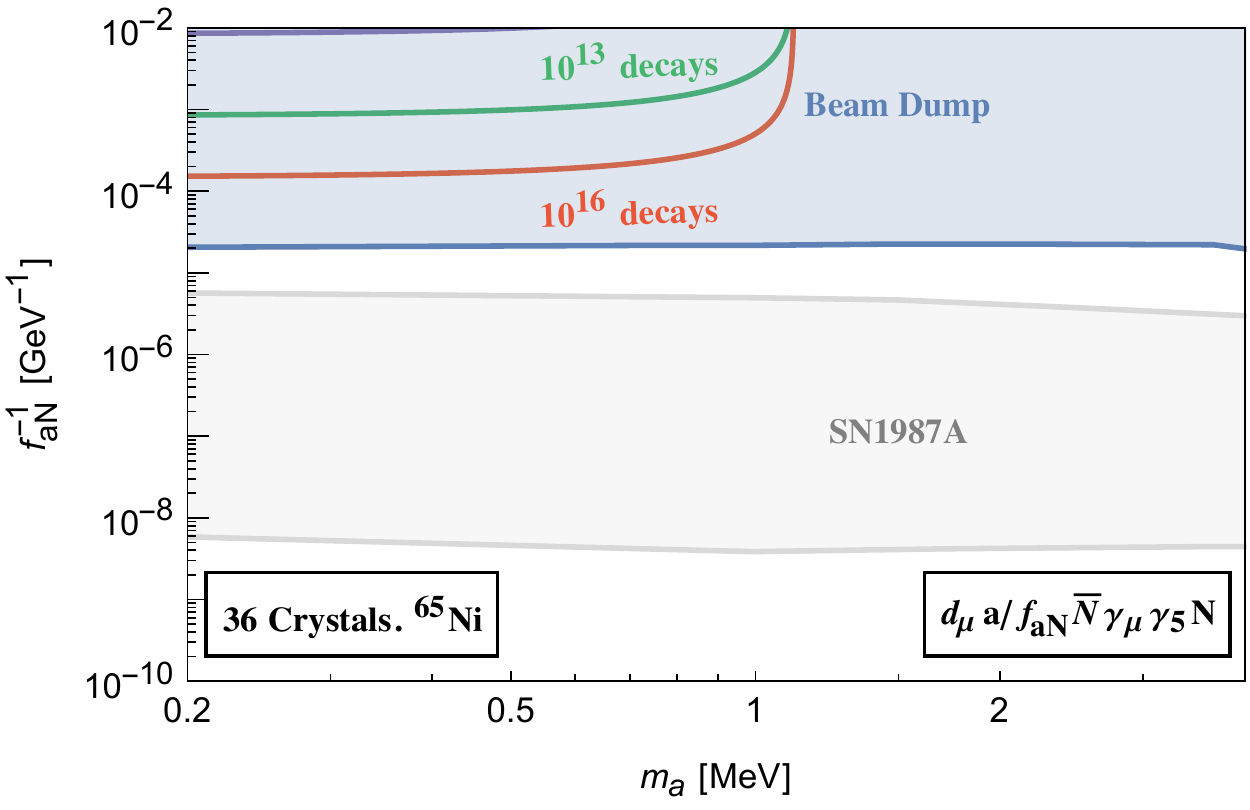} 
    \end{subfigure}
    \hfill
    \begin{subfigure}[t]{0.48\textwidth}
        \centering
        \includegraphics[width=\linewidth]{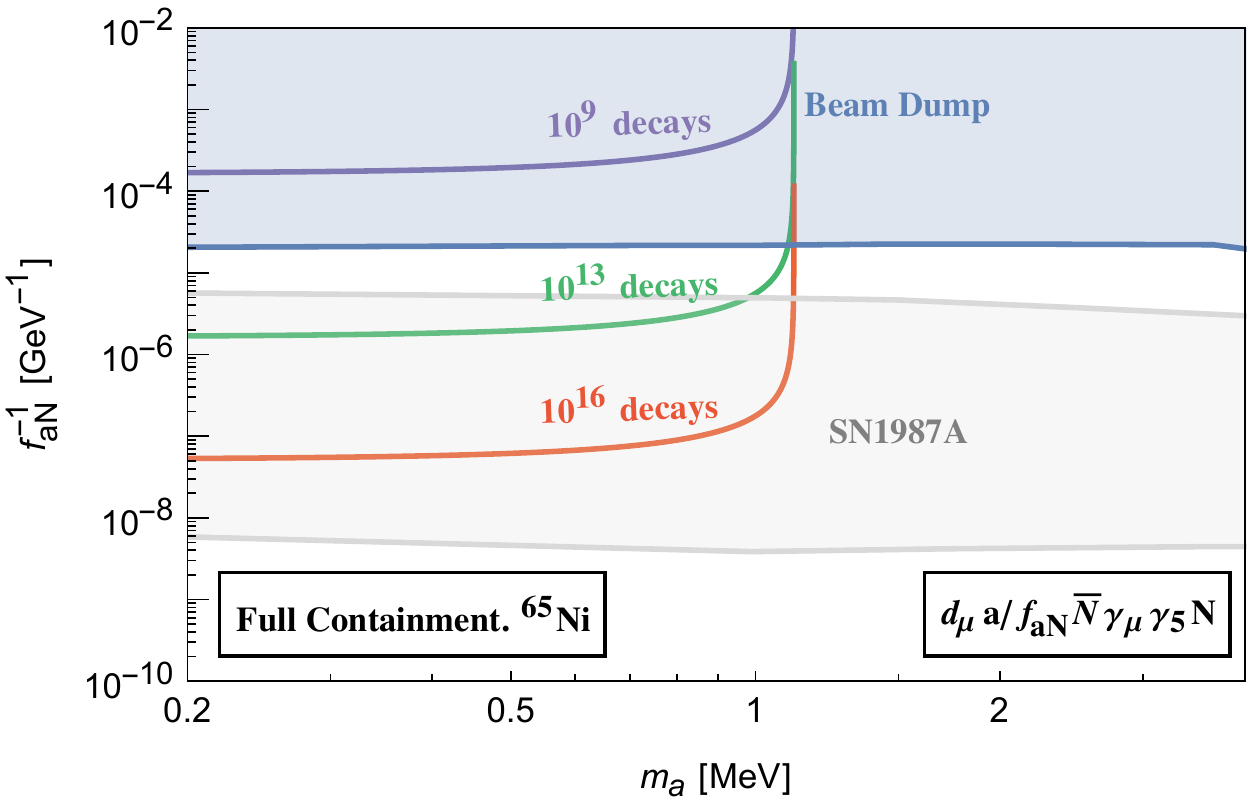} 
    \end{subfigure}
    
    \begin{subfigure}[t]{0.48\textwidth}
        \centering
        \includegraphics[width=\linewidth]{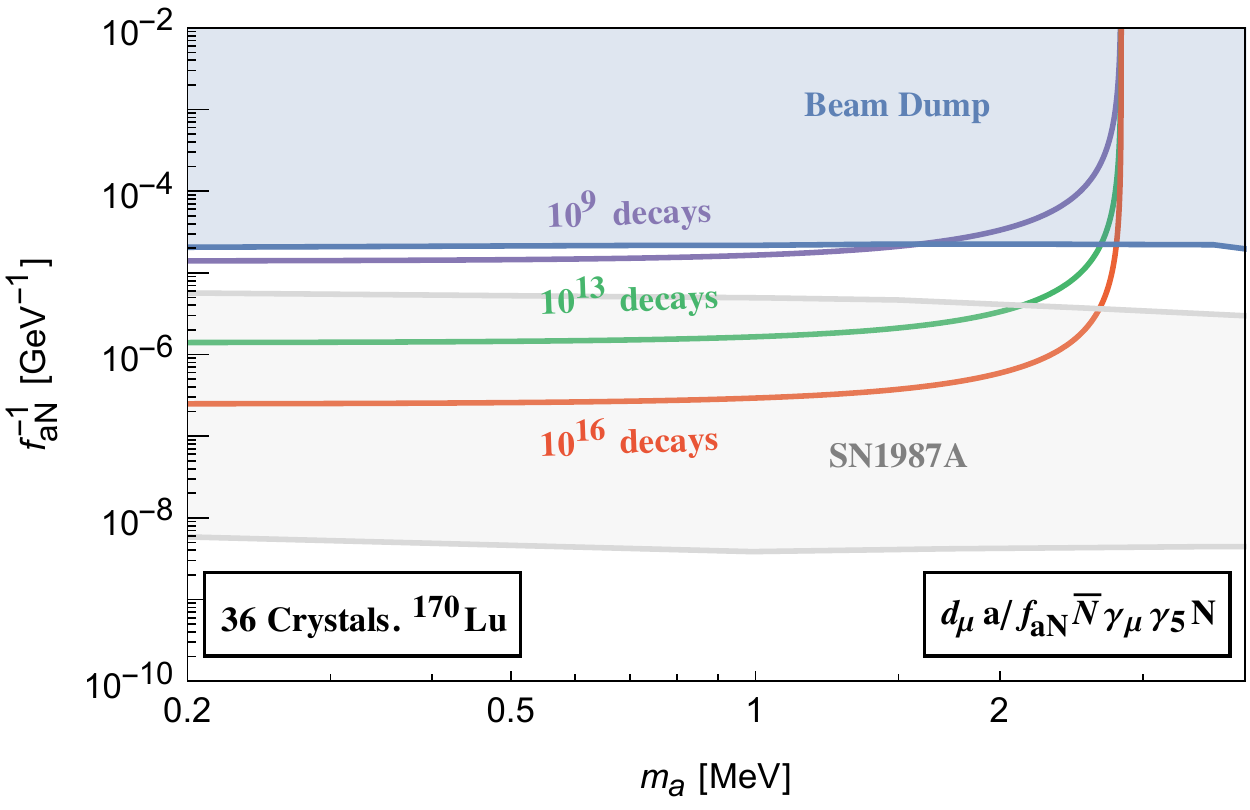} 
    \end{subfigure}
    \hfill
    \begin{subfigure}[t]{0.48\textwidth}
        \centering
        \includegraphics[width=\linewidth]{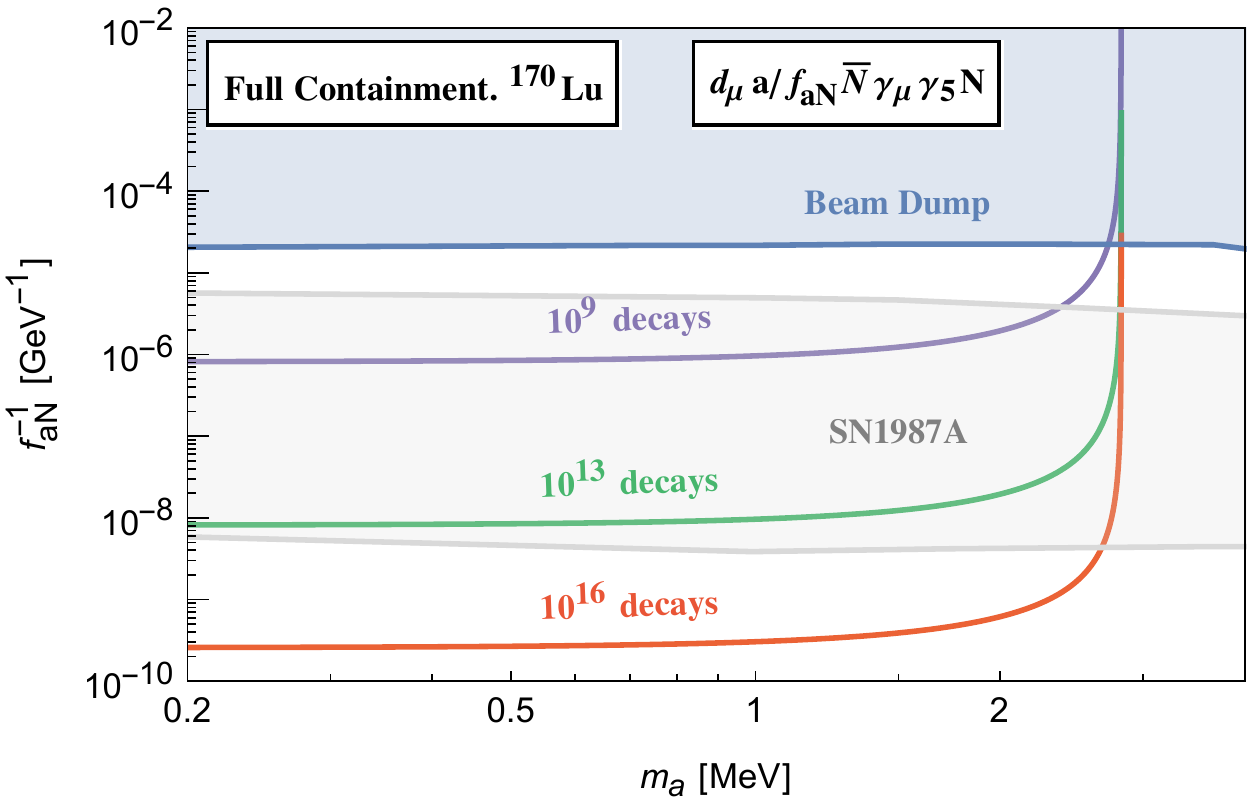} 
    \end{subfigure}
     \caption{Reach for ALP coupled to protons for $M_1$ (\textbf{top}) and $M_0$  (\textbf{bottom}) transitions for  current (\textbf{left panel}) and future (\textbf{right panel}) scintillator configurations.}
     \label{fig:alp_nuclear_limits}
\end{figure}

The projected reach for millicharge particles with charge $Q$ and mass $m_Q$ is shown in Fig.~\ref{fig:millicharge_limits}. Existing laboratory limits arise from the SLAC millicharge experiment \cite{Prinz:1998ua}, while astrophysical and cosmological limits are obtained from \cite{Chang:2018rso, Vogel:2013raa, Berlin:2020pey}. A $^{90}$Nb source with full containment performs excellently here as well. 
In Fig.~\ref{fig:alp_nuclear_limits}, we show projections for the ALP-nucleon coupling, with $M_1$ transitions in the decay chain of $^{65}$Ni in the top row and $M_0$ transitions in the decay chain of $^{170}$Lu in the bottom row. Existing limits arise from SN1987a~\cite{Chang:2018rso} and beam dumps~\cite{Ramani:2019jam}. New parameter space can be probed with $10^9$ decays even in the current setup with $^{170}$Lu nuclei. With full containment, parameter space below the supernova cooling constraints can be reached with $^{170}$Lu owing to the favorable selection rules which enhance the ALP decay branching fraction.

\begin{figure}[h]
    \centering
    \begin{subfigure}[t]{0.48\textwidth}
        \centering
        \includegraphics[width=\linewidth]{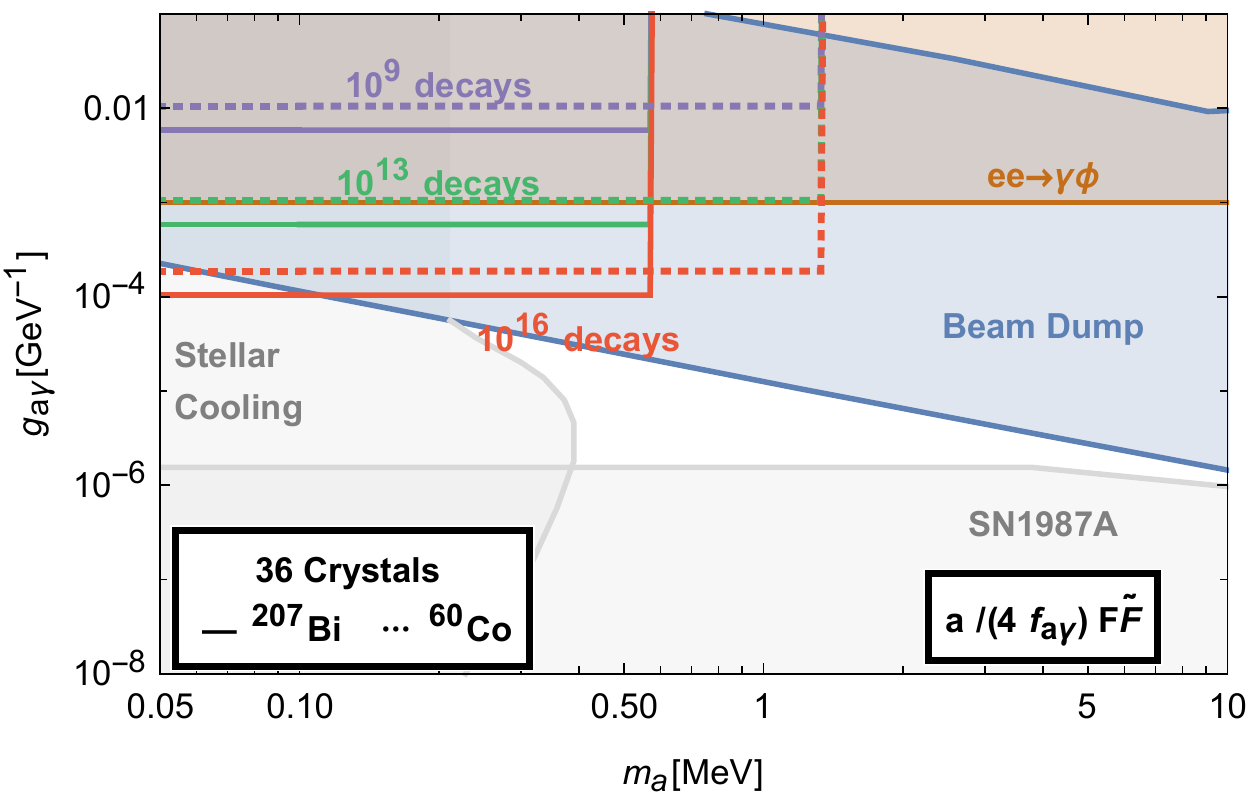} 
    \end{subfigure}
    \hfill
    \begin{subfigure}[t]{0.48\textwidth}
        \centering
        \includegraphics[width=\linewidth]{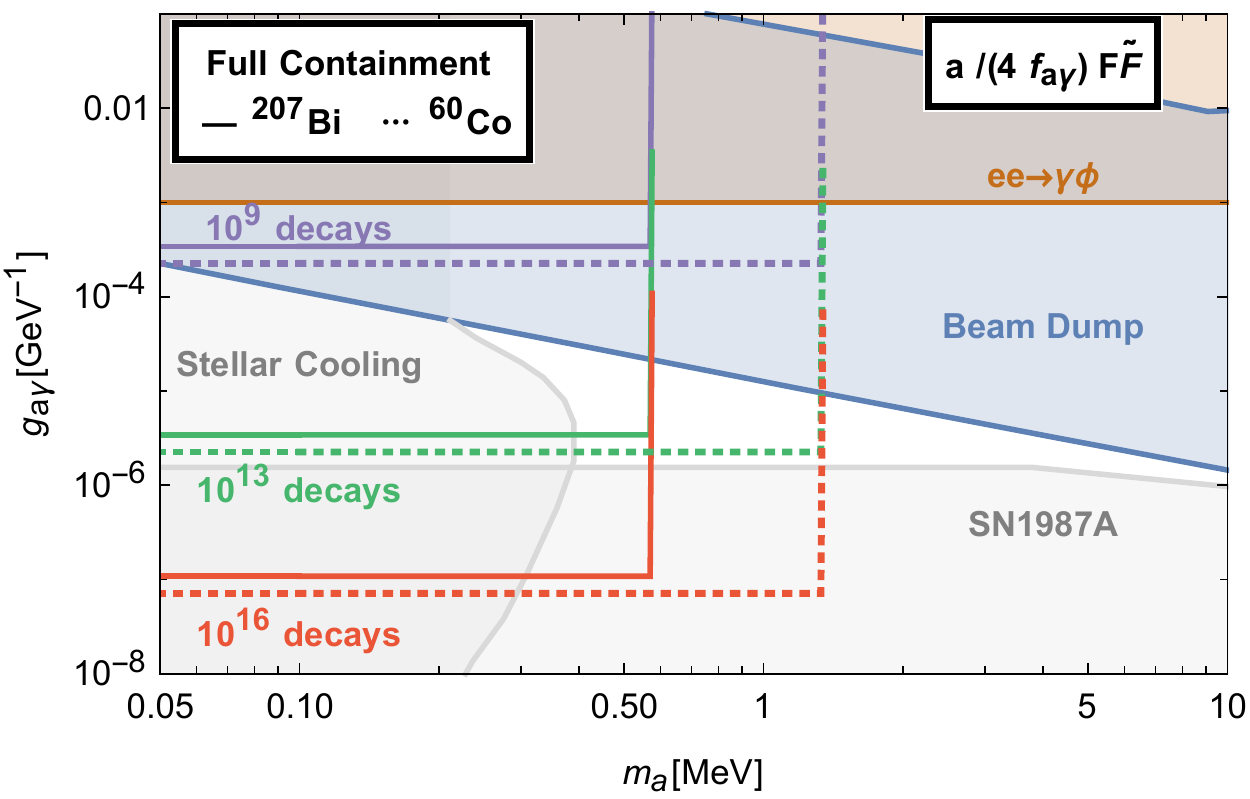} 
    \end{subfigure}
     \caption{Reach for ALPs coupled to photons for $E_2$ photons converting to axions via Primakoff scattering in the detector for current (\textbf{left panel}) and future (\textbf{right panel}) scintillator configurations. }     \label{fig:alp_gamma_limits}
\end{figure}

\begin{figure}[h]
    \centering
    \begin{subfigure}[t]{0.48\textwidth}
        \centering
        \includegraphics[width=\linewidth]{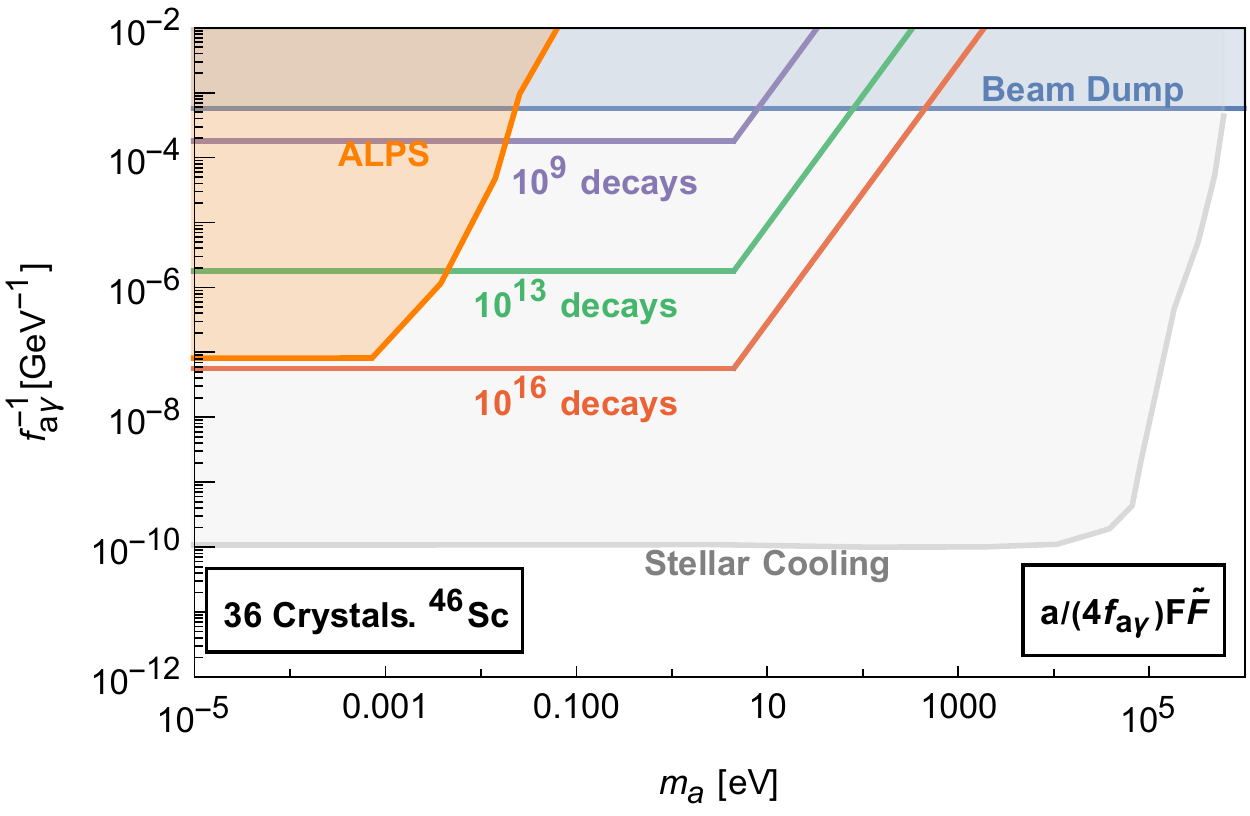} 
    \end{subfigure}
    \hfill
    \begin{subfigure}[t]{0.48\textwidth}
        \centering
        \includegraphics[width=\linewidth]{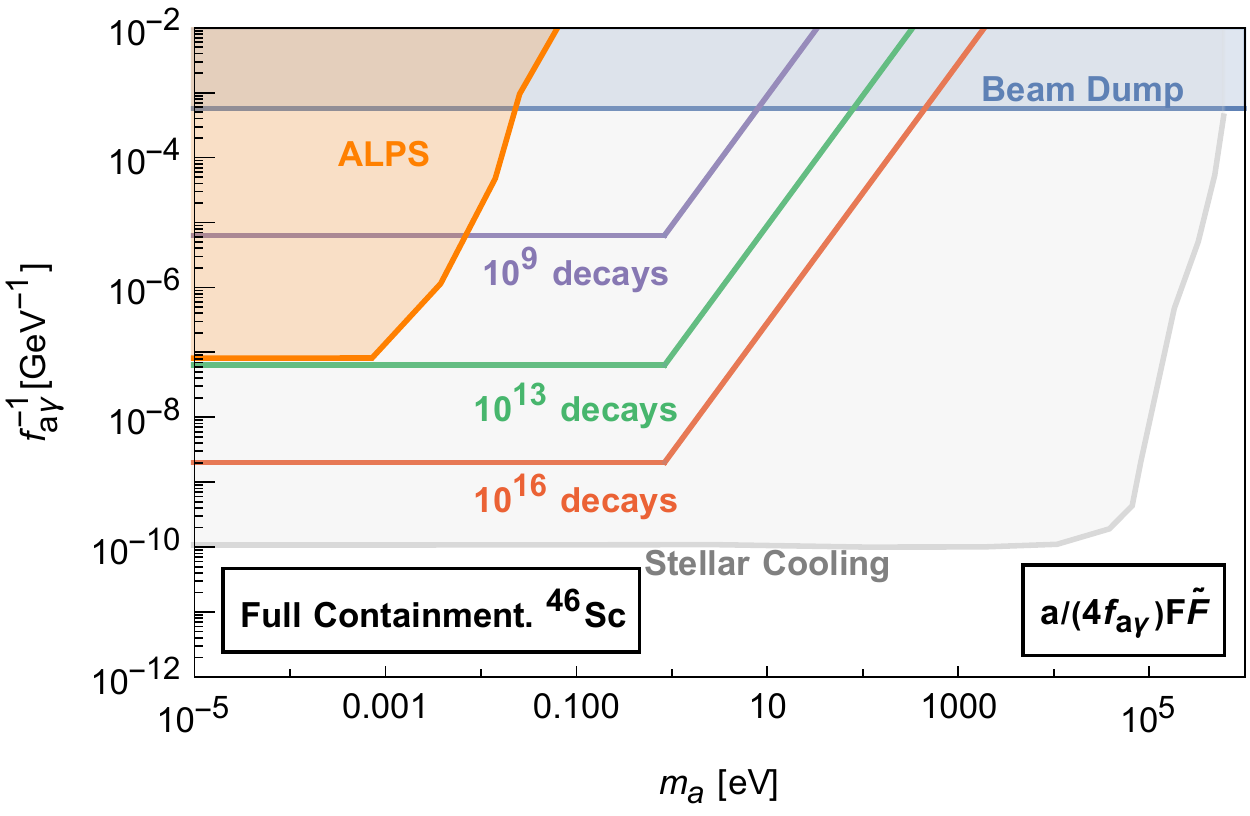} 
    \end{subfigure}
     \caption{Reach for ALPs coupled to photons for $E_2$ photons converting to axions in the presence of an external magnetic field $B_{\rm ext}=1$T for current (\textbf{left panel}) and future (\textbf{right panel}) scintillator configurations.}
     \label{fig:alp_gamma_mag}
\end{figure}

We finally discuss sensitivity of ALP coupling to photons. 
The projected sensitivity via Primakoff conversion in the detector material is shown in Fig.~\ref{fig:alp_gamma_limits} for both $^{207}$Bi and $^{60}$Co sources. We show both sources to illustrate that while $^{60}$Co has a wider range of $m_a$ sensitivity ($E_\text{probe}=1.33$~MeV), a $^{207}$Bi source has deeper reach due to the higher absorption of the $^{207}$Bi probe gamma ($E_\text{probe} = 0.57$ MeV). However, existing constraints from beam dump experiments~\cite{Bauer:2017ris} limit the reach over new parameter space to a full-containment scenario ($\sim 400$ crystals) with a larger activity exposure ($\gtrsim 10^{16}$ decays). With full-containment, the reach can cover the ``cosmological triangle"~\cite{Carenza:2020zil,Depta:2020wmr}, which is still allowed by the astrophysical and beam dump data.   In Fig.~\ref{fig:alp_gamma_mag}, we also show the effect of using a magnetic field to induce the coherent conversion of photons over the entire spatial extent of the field instead of using atomic conversion in material. Due to the coherence condition in Eq.~\eqref{eq:mag_primakoff}, however, this does limit the reach to lighter ALP masses. Although existing constraints cover the sensitive area of parameter space, a large part of this comes from stellar cooling (HB stars) which are model-dependent as discussed above. 

Lastly, new parameter space for direct ALP-electron couplings can be probed. Like in the $g_{a\gamma}$ case, this sensitivity is driven by photon-ALP conversion in material, but this time through the Compton-like conversion $\gamma \, e^- \rightarrow a \, e^-$. The reach with full containment in such a scenario can be $g_{ae} \sim 10^{-4}$ for $m_a < 0.75$ MeV. Since this may not be competitive with existing laboratory limits, we omit the projections for this work. However, a future experimental setup with a more active source may be able to expand the sensitivity in this parameter space.

\section{Conclusions}

The investigations performed in this paper using the 36 Cs(Tl) scintillators available at Texas A\&M show that a proof of concept experimental apparatus using these scintillators can yield science results in nuclei with $E_0$ and $M_0$ transitions, probing presently unconstrained parameter space for scalars and pseudoscalars coupled to nucleons. The key limitation of this  apparatus for other multipole transitions is containment - with 36 crystals, unlike the $E_0$ and $M_0$ transitions that produce difficult to miss electrons and positrons, the gammas produced in other transitions can be missed if the detector is not sufficiently big. A successful demonstration of this proof of concept setup would sharpen the case for the construction of a larger detector that is able to probe this broader class of transitions. These investigations also support the case for  modular development of this experiment with an inner module that is made of crystal scintillators and an outer module that could be constructed from liquid or plastic scintillators. This is because of the fact that the required energy resolution requires shorter scintillating modules to avoid loss of photons from absorption. The inner modules of the detector, where the vast majority of the produced gammas will be absorbed, thus need to be well instrumented. The outer modules, which constitute the bulk of the volume of the detector, exist to observe rare gammas that exit the inner module without interactions. Due to the low probability associated with this possibility, the requirements on energy resolution in the outer modules is more relaxed, enlarging the range of detector material.  This modular approach is thus likely to be cost effective without sacrificing the science potential of the experiment. 

The strong science case for laboratory probes of light, weakly coupled particles warrants investigation of other technologies that could be used to search for these particles using this protocol. For example, investments in dark matter detection has resulted in the development of sensors that can detect eV scale nuclear recoils with low background and excellent rejection between nuclear and electron recoils. It would be interesting to incorporate this technology into this experiment - with eV scale detection, the nuclear recoil produced from a decay can be directly detected without the need for a gamma cascade, enlarging the range of nuclei that could be used for this kind of detection concept. In fact, this technology could be used to potentially probe missing energy in decays involving electron capture - the nuclear recoil from the electron capture can be observed and the experiment can search for the missing Auger electrons/photons, both of which are difficult to miss. This kind of experiment would be uniquely sensitive to new physics coupled to electrons, complementing the probes of particles that couple to nucleons and photons presented here.  

\begin{acknowledgments}
The work of RM, AJ and SV are supported by DOE Grant Nos DE-SC0018981, DE-SC0017859, and DE-SC0021051, as well as internal funding from the Mitchell Institute. JBD acknowledges support from the U.S. National Science Foundation under Grant No. NSF PHY-1820801. The work of BD, DK and AT are supported  by the DOE Grant No.~DE-SC0010813. SR is supported by  the NSF under grant PHY-1818899, the SQMS Quantum Center and DoE support for MAGIS. RM acknowledges the critical contribution of the CsI crystals by Jim Alexander from the CLEO experiment at Cornell University. HR is supported in part by NSF Grant PHY-1720397 and the Gordon and Betty Moore Foundation Grant GBMF7946.
\end{acknowledgments}


\clearpage
\bibliography{sample}

\begin{thebibliography}{48}
\expandafter\ifx\csname natexlab\endcsname\relax\def\natexlab#1{#1}\fi
\expandafter\ifx\csname bibnamefont\endcsname\relax
  \def\bibnamefont#1{#1}\fi
\expandafter\ifx\csname bibfnamefont\endcsname\relax
  \def\bibfnamefont#1{#1}\fi
\expandafter\ifx\csname citenamefont\endcsname\relax
  \def\citenamefont#1{#1}\fi
\expandafter\ifx\csname url\endcsname\relax
  \def\url#1{\texttt{#1}}\fi
\expandafter\ifx\csname urlprefix\endcsname\relax\def\urlprefix{URL }\fi
\providecommand{\bibinfo}[2]{#2}
\providecommand{\eprint}[2][]{\url{#2}}

\bibitem[{\citenamefont{Green and Rajendran}(2017)}]{Green:2017ybv}
\bibinfo{author}{\bibfnamefont{D.}~\bibnamefont{Green}} \bibnamefont{and}
  \bibinfo{author}{\bibfnamefont{S.}~\bibnamefont{Rajendran}},
  \bibinfo{journal}{JHEP} \textbf{\bibinfo{volume}{10}}, \bibinfo{pages}{013}
  (\bibinfo{year}{2017}), \eprint{1701.08750}.

\bibitem[{\citenamefont{Peccei and Quinn}(1977)}]{Peccei:1977hh}
\bibinfo{author}{\bibfnamefont{R.~D.} \bibnamefont{Peccei}} \bibnamefont{and}
  \bibinfo{author}{\bibfnamefont{H.~R.} \bibnamefont{Quinn}},
  \bibinfo{journal}{Phys. Rev. Lett.} \textbf{\bibinfo{volume}{38}},
  \bibinfo{pages}{1440} (\bibinfo{year}{1977}).

\bibitem[{\citenamefont{Weinberg}(1978)}]{Weinberg:1977ma}
\bibinfo{author}{\bibfnamefont{S.}~\bibnamefont{Weinberg}},
  \bibinfo{journal}{Phys. Rev. Lett.} \textbf{\bibinfo{volume}{40}},
  \bibinfo{pages}{223} (\bibinfo{year}{1978}).

\bibitem[{\citenamefont{Wilczek}(1978)}]{Wilczek:1977pj}
\bibinfo{author}{\bibfnamefont{F.}~\bibnamefont{Wilczek}},
  \bibinfo{journal}{Phys. Rev. Lett.} \textbf{\bibinfo{volume}{40}},
  \bibinfo{pages}{279} (\bibinfo{year}{1978}).

\bibitem[{\citenamefont{Graham et~al.}(2015)\citenamefont{Graham, Kaplan, and
  Rajendran}}]{Graham:2015cka}
\bibinfo{author}{\bibfnamefont{P.~W.} \bibnamefont{Graham}},
  \bibinfo{author}{\bibfnamefont{D.~E.} \bibnamefont{Kaplan}},
  \bibnamefont{and}
  \bibinfo{author}{\bibfnamefont{S.}~\bibnamefont{Rajendran}},
  \bibinfo{journal}{Phys. Rev. Lett.} \textbf{\bibinfo{volume}{115}},
  \bibinfo{pages}{221801} (\bibinfo{year}{2015}), \eprint{1504.07551}.

\bibitem[{\citenamefont{Graham et~al.}(2018)\citenamefont{Graham, Kaplan, and
  Rajendran}}]{Graham:2017hfr}
\bibinfo{author}{\bibfnamefont{P.~W.} \bibnamefont{Graham}},
  \bibinfo{author}{\bibfnamefont{D.~E.} \bibnamefont{Kaplan}},
  \bibnamefont{and}
  \bibinfo{author}{\bibfnamefont{S.}~\bibnamefont{Rajendran}},
  \bibinfo{journal}{Phys. Rev. D} \textbf{\bibinfo{volume}{97}},
  \bibinfo{pages}{044003} (\bibinfo{year}{2018}), \eprint{1709.01999}.

\bibitem[{\citenamefont{Graham et~al.}(2019)\citenamefont{Graham, Kaplan, and
  Rajendran}}]{Graham:2019bfu}
\bibinfo{author}{\bibfnamefont{P.~W.} \bibnamefont{Graham}},
  \bibinfo{author}{\bibfnamefont{D.~E.} \bibnamefont{Kaplan}},
  \bibnamefont{and}
  \bibinfo{author}{\bibfnamefont{S.}~\bibnamefont{Rajendran}},
  \bibinfo{journal}{Phys. Rev. D} \textbf{\bibinfo{volume}{100}},
  \bibinfo{pages}{015048} (\bibinfo{year}{2019}), \eprint{1902.06793}.

\bibitem[{\citenamefont{DeRocco et~al.}(2020)\citenamefont{DeRocco, Graham, and
  Rajendran}}]{DeRocco:2020xdt}
\bibinfo{author}{\bibfnamefont{W.}~\bibnamefont{DeRocco}},
  \bibinfo{author}{\bibfnamefont{P.~W.} \bibnamefont{Graham}},
  \bibnamefont{and}
  \bibinfo{author}{\bibfnamefont{S.}~\bibnamefont{Rajendran}},
  \bibinfo{journal}{Phys. Rev. D} \textbf{\bibinfo{volume}{102}},
  \bibinfo{pages}{075015} (\bibinfo{year}{2020}), \eprint{2006.15112}.

\bibitem[{\citenamefont{Benato et~al.}(2019)\citenamefont{Benato, Drobizhev,
  Rajendran, and Ramani}}]{Benato:2018ijc}
\bibinfo{author}{\bibfnamefont{G.}~\bibnamefont{Benato}},
  \bibinfo{author}{\bibfnamefont{A.}~\bibnamefont{Drobizhev}},
  \bibinfo{author}{\bibfnamefont{S.}~\bibnamefont{Rajendran}},
  \bibnamefont{and} \bibinfo{author}{\bibfnamefont{H.}~\bibnamefont{Ramani}},
  \bibinfo{journal}{Phys. Rev.} \textbf{\bibinfo{volume}{D99}},
  \bibinfo{pages}{035025} (\bibinfo{year}{2019}), \eprint{1810.06467}.

\bibitem[{\citenamefont{Blucher et~al.}(1986)\citenamefont{Blucher, Gittelman,
  Heltsley, Kandaswamy, Kowalewski, Kubota, Mistry, Stone, and
  Bean}}]{Blucher:1986rv}
\bibinfo{author}{\bibfnamefont{E.}~\bibnamefont{Blucher}},
  \bibinfo{author}{\bibfnamefont{B.}~\bibnamefont{Gittelman}},
  \bibinfo{author}{\bibfnamefont{B.}~\bibnamefont{Heltsley}},
  \bibinfo{author}{\bibfnamefont{J.}~\bibnamefont{Kandaswamy}},
  \bibinfo{author}{\bibfnamefont{R.}~\bibnamefont{Kowalewski}},
  \bibinfo{author}{\bibfnamefont{Y.}~\bibnamefont{Kubota}},
  \bibinfo{author}{\bibfnamefont{N.}~\bibnamefont{Mistry}},
  \bibinfo{author}{\bibfnamefont{S.}~\bibnamefont{Stone}}, \bibnamefont{and}
  \bibinfo{author}{\bibfnamefont{A.}~\bibnamefont{Bean}},
  \bibinfo{journal}{Nuclear Instruments and Methods in Physics Research Section
  A} \textbf{\bibinfo{volume}{249}}, \bibinfo{pages}{201}
  (\bibinfo{year}{1986}).

\bibitem[{\citenamefont{Brun et~al.}(2003)\citenamefont{Brun, Gheata, and
  Gheata}}]{Brun:2003xr}
\bibinfo{author}{\bibfnamefont{R.}~\bibnamefont{Brun}},
  \bibinfo{author}{\bibfnamefont{A.}~\bibnamefont{Gheata}}, \bibnamefont{and}
  \bibinfo{author}{\bibfnamefont{M.}~\bibnamefont{Gheata}},
  \bibinfo{journal}{Nucl. Instrum. Meth. A} \textbf{\bibinfo{volume}{502}},
  \bibinfo{pages}{676} (\bibinfo{year}{2003}).

\bibitem[{\citenamefont{Agostinelli et~al.}(2003)}]{Agostinelli:2002hh}
\bibinfo{author}{\bibfnamefont{S.}~\bibnamefont{Agostinelli}}
  \bibnamefont{et~al.} (\bibinfo{collaboration}{GEANT4}),
  \bibinfo{journal}{Nucl. Instrum. Meth. A} \textbf{\bibinfo{volume}{506}},
  \bibinfo{pages}{250} (\bibinfo{year}{2003}).

\bibitem[{\citenamefont{Avignone et~al.}(1988)\citenamefont{Avignone, Baktash,
  Barker, Calaprice, Dunford, Haxton, Kahana, Kouzes, Miley, and
  Moltz}}]{Avignone:1988bv}
\bibinfo{author}{\bibfnamefont{F.~T.} \bibnamefont{Avignone}},
  \bibinfo{author}{\bibfnamefont{C.}~\bibnamefont{Baktash}},
  \bibinfo{author}{\bibfnamefont{W.~C.} \bibnamefont{Barker}},
  \bibinfo{author}{\bibfnamefont{F.~P.} \bibnamefont{Calaprice}},
  \bibinfo{author}{\bibfnamefont{R.~W.} \bibnamefont{Dunford}},
  \bibinfo{author}{\bibfnamefont{W.~C.} \bibnamefont{Haxton}},
  \bibinfo{author}{\bibfnamefont{D.}~\bibnamefont{Kahana}},
  \bibinfo{author}{\bibfnamefont{R.~T.} \bibnamefont{Kouzes}},
  \bibinfo{author}{\bibfnamefont{H.~S.} \bibnamefont{Miley}}, \bibnamefont{and}
  \bibinfo{author}{\bibfnamefont{D.~M.} \bibnamefont{Moltz}},
  \bibinfo{journal}{Phys. Rev. D} \textbf{\bibinfo{volume}{37}},
  \bibinfo{pages}{618} (\bibinfo{year}{1988}).

\bibitem[{\citenamefont{Pettersson et~al.}(1968)\citenamefont{Pettersson,
  Antman, and Grunditz}}]{pettersson1968decay}
\bibinfo{author}{\bibfnamefont{H.}~\bibnamefont{Pettersson}},
  \bibinfo{author}{\bibfnamefont{S.}~\bibnamefont{Antman}}, \bibnamefont{and}
  \bibinfo{author}{\bibfnamefont{Y.}~\bibnamefont{Grunditz}},
  \bibinfo{journal}{Nuclear Physics A} \textbf{\bibinfo{volume}{108}},
  \bibinfo{pages}{124} (\bibinfo{year}{1968}).

\bibitem[{\citenamefont{Warburton and Alburger}(1982)}]{warburton1982decay}
\bibinfo{author}{\bibfnamefont{E.}~\bibnamefont{Warburton}} \bibnamefont{and}
  \bibinfo{author}{\bibfnamefont{D.}~\bibnamefont{Alburger}},
  \bibinfo{journal}{Physical Review. C, Nuclear Physics}
  \textbf{\bibinfo{volume}{26}}, \bibinfo{pages}{2595} (\bibinfo{year}{1982}).

\bibitem[{\citenamefont{D'Arienzo}(2013)}]{d2013emission}
\bibinfo{author}{\bibfnamefont{M.}~\bibnamefont{D'Arienzo}},
  \bibinfo{journal}{Atoms} \textbf{\bibinfo{volume}{1}}, \bibinfo{pages}{2}
  (\bibinfo{year}{2013}).

\bibitem[{\citenamefont{Schirmer et~al.}(1984)\citenamefont{Schirmer, Habs,
  Kroth, Kwong, Schwalm, Zirnbauer, and Broude}}]{schirmer1984double}
\bibinfo{author}{\bibfnamefont{J.}~\bibnamefont{Schirmer}},
  \bibinfo{author}{\bibfnamefont{D.}~\bibnamefont{Habs}},
  \bibinfo{author}{\bibfnamefont{R.}~\bibnamefont{Kroth}},
  \bibinfo{author}{\bibfnamefont{N.}~\bibnamefont{Kwong}},
  \bibinfo{author}{\bibfnamefont{D.}~\bibnamefont{Schwalm}},
  \bibinfo{author}{\bibfnamefont{M.}~\bibnamefont{Zirnbauer}},
  \bibnamefont{and} \bibinfo{author}{\bibfnamefont{C.}~\bibnamefont{Broude}},
  \bibinfo{journal}{Physical Review Letters} \textbf{\bibinfo{volume}{53}},
  \bibinfo{pages}{1897} (\bibinfo{year}{1984}).

\bibitem[{\citenamefont{Krutov and Knyazkov}(1970)}]{krutov1970higher}
\bibinfo{author}{\bibfnamefont{V.}~\bibnamefont{Krutov}} \bibnamefont{and}
  \bibinfo{author}{\bibfnamefont{O.}~\bibnamefont{Knyazkov}},
  \bibinfo{journal}{Annalen der Physik} \textbf{\bibinfo{volume}{480}},
  \bibinfo{pages}{10} (\bibinfo{year}{1970}).

\bibitem[{\citenamefont{Krasznahorkay et~al.}()\citenamefont{Krasznahorkay,
  de~Boera, Guly{\'a}s, G{\'a}csi, Ketela, Csatl{\'o}s, Csige, Hunyadi, van
  Klinkenb, Krasznahorkay~Jr et~al.}}]{krasznahorkay2}
\bibinfo{author}{\bibfnamefont{A.}~\bibnamefont{Krasznahorkay}},
  \bibinfo{author}{\bibfnamefont{F.}~\bibnamefont{de~Boera}},
  \bibinfo{author}{\bibfnamefont{J.}~\bibnamefont{Guly{\'a}s}},
  \bibinfo{author}{\bibfnamefont{Z.}~\bibnamefont{G{\'a}csi}},
  \bibinfo{author}{\bibfnamefont{T.}~\bibnamefont{Ketela}},
  \bibinfo{author}{\bibfnamefont{M.}~\bibnamefont{Csatl{\'o}s}},
  \bibinfo{author}{\bibfnamefont{L.}~\bibnamefont{Csige}},
  \bibinfo{author}{\bibfnamefont{M.}~\bibnamefont{Hunyadi}},
  \bibinfo{author}{\bibfnamefont{J.}~\bibnamefont{van Klinkenb}},
  \bibinfo{author}{\bibfnamefont{A.}~\bibnamefont{Krasznahorkay~Jr}},
  \bibnamefont{et~al.} (????).

\bibitem[{\citenamefont{Alburger}(1978)}]{alburger1978comment}
\bibinfo{author}{\bibfnamefont{D.~E.} \bibnamefont{Alburger}},
  \bibinfo{journal}{Physical Review C} \textbf{\bibinfo{volume}{18}},
  \bibinfo{pages}{576} (\bibinfo{year}{1978}).

\bibitem[{\citenamefont{Kuhnert et~al.}(1993)\citenamefont{Kuhnert, Henry,
  Wang, Brinkman, Stoyer, Becker, Manatt, and Yates}}]{kuhnert1993search}
\bibinfo{author}{\bibfnamefont{A.}~\bibnamefont{Kuhnert}},
  \bibinfo{author}{\bibfnamefont{E.}~\bibnamefont{Henry}},
  \bibinfo{author}{\bibfnamefont{T.}~\bibnamefont{Wang}},
  \bibinfo{author}{\bibfnamefont{M.}~\bibnamefont{Brinkman}},
  \bibinfo{author}{\bibfnamefont{M.}~\bibnamefont{Stoyer}},
  \bibinfo{author}{\bibfnamefont{J.}~\bibnamefont{Becker}},
  \bibinfo{author}{\bibfnamefont{D.}~\bibnamefont{Manatt}}, \bibnamefont{and}
  \bibinfo{author}{\bibfnamefont{S.}~\bibnamefont{Yates}},
  \bibinfo{journal}{Physical Review C} \textbf{\bibinfo{volume}{47}},
  \bibinfo{pages}{2386} (\bibinfo{year}{1993}).

\bibitem[{\citenamefont{Knapen et~al.}(2017)\citenamefont{Knapen, Lin, and
  Zurek}}]{Knapen:2017xzo}
\bibinfo{author}{\bibfnamefont{S.}~\bibnamefont{Knapen}},
  \bibinfo{author}{\bibfnamefont{T.}~\bibnamefont{Lin}}, \bibnamefont{and}
  \bibinfo{author}{\bibfnamefont{K.~M.} \bibnamefont{Zurek}},
  \bibinfo{journal}{Phys. Rev. D} \textbf{\bibinfo{volume}{96}},
  \bibinfo{pages}{115021} (\bibinfo{year}{2017}), \eprint{1709.07882}.

\bibitem[{\citenamefont{Dev et~al.}(2020)\citenamefont{Dev, Mohapatra, and
  Zhang}}]{Dev:2020eam}
\bibinfo{author}{\bibfnamefont{P.~S.~B.} \bibnamefont{Dev}},
  \bibinfo{author}{\bibfnamefont{R.~N.} \bibnamefont{Mohapatra}},
  \bibnamefont{and} \bibinfo{author}{\bibfnamefont{Y.}~\bibnamefont{Zhang}},
  \bibinfo{journal}{JCAP} \textbf{\bibinfo{volume}{08}}, \bibinfo{pages}{003}
  (\bibinfo{year}{2020}), \bibinfo{note}{[Erratum: JCAP 11, E01 (2020)]},
  \eprint{2005.00490}.

\bibitem[{\citenamefont{Izaguirre et~al.}(2015)\citenamefont{Izaguirre,
  Krnjaic, and Pospelov}}]{Izaguirre:2014cza}
\bibinfo{author}{\bibfnamefont{E.}~\bibnamefont{Izaguirre}},
  \bibinfo{author}{\bibfnamefont{G.}~\bibnamefont{Krnjaic}}, \bibnamefont{and}
  \bibinfo{author}{\bibfnamefont{M.}~\bibnamefont{Pospelov}},
  \bibinfo{journal}{Phys. Lett.} \textbf{\bibinfo{volume}{B740}},
  \bibinfo{pages}{61} (\bibinfo{year}{2015}), \eprint{1405.4864}.

\bibitem[{\citenamefont{Snover and Hurd}(2003)}]{snover2003e+}
\bibinfo{author}{\bibfnamefont{K.}~\bibnamefont{Snover}} \bibnamefont{and}
  \bibinfo{author}{\bibfnamefont{A.}~\bibnamefont{Hurd}},
  \bibinfo{journal}{Physical Review C} \textbf{\bibinfo{volume}{67}},
  \bibinfo{pages}{055801} (\bibinfo{year}{2003}).

\bibitem[{\citenamefont{Avignone~III et~al.}(1988)\citenamefont{Avignone~III,
  Baktash, Barker, Calaprice, Dunford, Haxton, Kahana, Kouzes, Miley, and
  Moltz}}]{avignone1988search}
\bibinfo{author}{\bibfnamefont{F.}~\bibnamefont{Avignone~III}},
  \bibinfo{author}{\bibfnamefont{C.}~\bibnamefont{Baktash}},
  \bibinfo{author}{\bibfnamefont{W.}~\bibnamefont{Barker}},
  \bibinfo{author}{\bibfnamefont{F.}~\bibnamefont{Calaprice}},
  \bibinfo{author}{\bibfnamefont{R.}~\bibnamefont{Dunford}},
  \bibinfo{author}{\bibfnamefont{W.}~\bibnamefont{Haxton}},
  \bibinfo{author}{\bibfnamefont{D.}~\bibnamefont{Kahana}},
  \bibinfo{author}{\bibfnamefont{R.}~\bibnamefont{Kouzes}},
  \bibinfo{author}{\bibfnamefont{H.}~\bibnamefont{Miley}}, \bibnamefont{and}
  \bibinfo{author}{\bibfnamefont{D.}~\bibnamefont{Moltz}},
  \bibinfo{journal}{Physical Review D} \textbf{\bibinfo{volume}{37}},
  \bibinfo{pages}{618} (\bibinfo{year}{1988}).

\bibitem[{\citenamefont{Tsai}(1986)}]{Tsai:1986tx}
\bibinfo{author}{\bibfnamefont{Y.-S.} \bibnamefont{Tsai}},
  \bibinfo{journal}{Phys. Rev.} \textbf{\bibinfo{volume}{D34}},
  \bibinfo{pages}{1326} (\bibinfo{year}{1986}).

\bibitem[{\citenamefont{Raffelt}(1996)}]{Raffelt:1996wa}
\bibinfo{author}{\bibfnamefont{G.~G.} \bibnamefont{Raffelt}},
  \emph{\bibinfo{title}{{Stars as laboratories for fundamental physics}: {The
  astrophysics of neutrinos, axions, and other weakly interacting particles}}}
  (\bibinfo{year}{1996}), ISBN \bibinfo{isbn}{978-0-226-70272-8}.

\bibitem[{\citenamefont{Jaeckel et~al.}(2007)\citenamefont{Jaeckel, Masso,
  Redondo, Ringwald, and Takahashi}}]{Jaeckel:2006xm}
\bibinfo{author}{\bibfnamefont{J.}~\bibnamefont{Jaeckel}},
  \bibinfo{author}{\bibfnamefont{E.}~\bibnamefont{Masso}},
  \bibinfo{author}{\bibfnamefont{J.}~\bibnamefont{Redondo}},
  \bibinfo{author}{\bibfnamefont{A.}~\bibnamefont{Ringwald}}, \bibnamefont{and}
  \bibinfo{author}{\bibfnamefont{F.}~\bibnamefont{Takahashi}},
  \bibinfo{journal}{Phys. Rev. D} \textbf{\bibinfo{volume}{75}},
  \bibinfo{pages}{013004} (\bibinfo{year}{2007}), \eprint{hep-ph/0610203}.

\bibitem[{\citenamefont{Khoury and Weltman}(2004)}]{Khoury:2003aq}
\bibinfo{author}{\bibfnamefont{J.}~\bibnamefont{Khoury}} \bibnamefont{and}
  \bibinfo{author}{\bibfnamefont{A.}~\bibnamefont{Weltman}},
  \bibinfo{journal}{Phys. Rev. Lett.} \textbf{\bibinfo{volume}{93}},
  \bibinfo{pages}{171104} (\bibinfo{year}{2004}), \eprint{astro-ph/0309300}.

\bibitem[{\citenamefont{Masso and Redondo}(2005)}]{Masso:2005ym}
\bibinfo{author}{\bibfnamefont{E.}~\bibnamefont{Masso}} \bibnamefont{and}
  \bibinfo{author}{\bibfnamefont{J.}~\bibnamefont{Redondo}},
  \bibinfo{journal}{JCAP} \textbf{\bibinfo{volume}{0509}}, \bibinfo{pages}{015}
  (\bibinfo{year}{2005}), \eprint{hep-ph/0504202}.

\bibitem[{\citenamefont{Masso and Redondo}(2006)}]{Masso:2006gc}
\bibinfo{author}{\bibfnamefont{E.}~\bibnamefont{Masso}} \bibnamefont{and}
  \bibinfo{author}{\bibfnamefont{J.}~\bibnamefont{Redondo}},
  \bibinfo{journal}{Phys. Rev. Lett.} \textbf{\bibinfo{volume}{97}},
  \bibinfo{pages}{151802} (\bibinfo{year}{2006}), \eprint{hep-ph/0606163}.

\bibitem[{\citenamefont{Dupays et~al.}(2007)\citenamefont{Dupays, Masso,
  Redondo, and Rizzo}}]{Dupays:2006dp}
\bibinfo{author}{\bibfnamefont{A.}~\bibnamefont{Dupays}},
  \bibinfo{author}{\bibfnamefont{E.}~\bibnamefont{Masso}},
  \bibinfo{author}{\bibfnamefont{J.}~\bibnamefont{Redondo}}, \bibnamefont{and}
  \bibinfo{author}{\bibfnamefont{C.}~\bibnamefont{Rizzo}},
  \bibinfo{journal}{Phys. Rev. Lett.} \textbf{\bibinfo{volume}{98}},
  \bibinfo{pages}{131802} (\bibinfo{year}{2007}), \eprint{hep-ph/0610286}.

\bibitem[{\citenamefont{Mohapatra and Nasri}(2007)}]{Mohapatra:2006pv}
\bibinfo{author}{\bibfnamefont{R.~N.} \bibnamefont{Mohapatra}}
  \bibnamefont{and} \bibinfo{author}{\bibfnamefont{S.}~\bibnamefont{Nasri}},
  \bibinfo{journal}{Phys. Rev. Lett.} \textbf{\bibinfo{volume}{98}},
  \bibinfo{pages}{050402} (\bibinfo{year}{2007}), \eprint{hep-ph/0610068}.

\bibitem[{\citenamefont{Brax et~al.}(2007)\citenamefont{Brax, van~de Bruck, and
  Davis}}]{Brax:2007ak}
\bibinfo{author}{\bibfnamefont{P.}~\bibnamefont{Brax}},
  \bibinfo{author}{\bibfnamefont{C.}~\bibnamefont{van~de Bruck}},
  \bibnamefont{and} \bibinfo{author}{\bibfnamefont{A.-C.} \bibnamefont{Davis}},
  \bibinfo{journal}{Phys. Rev. Lett.} \textbf{\bibinfo{volume}{99}},
  \bibinfo{pages}{121103} (\bibinfo{year}{2007}), \eprint{hep-ph/0703243}.

\bibitem[{\citenamefont{Berlin and Hook}(2020)}]{Berlin:2020pey}
\bibinfo{author}{\bibfnamefont{A.}~\bibnamefont{Berlin}} \bibnamefont{and}
  \bibinfo{author}{\bibfnamefont{A.}~\bibnamefont{Hook}},
  \bibinfo{journal}{Phys. Rev. D} \textbf{\bibinfo{volume}{102}},
  \bibinfo{pages}{035010} (\bibinfo{year}{2020}), \eprint{2001.02679}.

\bibitem[{\citenamefont{Liu et~al.}(2016)\citenamefont{Liu, McKeen, and
  Miller}}]{Liu:2016qwd}
\bibinfo{author}{\bibfnamefont{Y.-S.} \bibnamefont{Liu}},
  \bibinfo{author}{\bibfnamefont{D.}~\bibnamefont{McKeen}}, \bibnamefont{and}
  \bibinfo{author}{\bibfnamefont{G.~A.} \bibnamefont{Miller}},
  \bibinfo{journal}{Phys. Rev. Lett.} \textbf{\bibinfo{volume}{117}},
  \bibinfo{pages}{101801} (\bibinfo{year}{2016}), \eprint{1605.04612}.

\bibitem[{\citenamefont{Artamonov et~al.}(2008)}]{Artamonov:2008qb}
\bibinfo{author}{\bibfnamefont{A.~V.} \bibnamefont{Artamonov}}
  \bibnamefont{et~al.} (\bibinfo{collaboration}{E949}), \bibinfo{journal}{Phys.
  Rev. Lett.} \textbf{\bibinfo{volume}{101}}, \bibinfo{pages}{191802}
  (\bibinfo{year}{2008}), \eprint{0808.2459}.

\bibitem[{\citenamefont{Banerjee et~al.}(2019)}]{NA64:2019imj}
\bibinfo{author}{\bibfnamefont{D.}~\bibnamefont{Banerjee}}
  \bibnamefont{et~al.}, \bibinfo{journal}{Phys. Rev. Lett.}
  \textbf{\bibinfo{volume}{123}}, \bibinfo{pages}{121801}
  (\bibinfo{year}{2019}), \eprint{1906.00176}.

\bibitem[{\citenamefont{Chang et~al.}(2017)\citenamefont{Chang, Essig, and
  McDermott}}]{Chang:2016ntp}
\bibinfo{author}{\bibfnamefont{J.~H.} \bibnamefont{Chang}},
  \bibinfo{author}{\bibfnamefont{R.}~\bibnamefont{Essig}}, \bibnamefont{and}
  \bibinfo{author}{\bibfnamefont{S.~D.} \bibnamefont{McDermott}},
  \bibinfo{journal}{JHEP} \textbf{\bibinfo{volume}{01}}, \bibinfo{pages}{107}
  (\bibinfo{year}{2017}), \eprint{1611.03864}.

\bibitem[{\citenamefont{Hardy and Lasenby}(2017)}]{Hardy:2016kme}
\bibinfo{author}{\bibfnamefont{E.}~\bibnamefont{Hardy}} \bibnamefont{and}
  \bibinfo{author}{\bibfnamefont{R.}~\bibnamefont{Lasenby}},
  \bibinfo{journal}{JHEP} \textbf{\bibinfo{volume}{02}}, \bibinfo{pages}{033}
  (\bibinfo{year}{2017}), \eprint{1611.05852}.

\bibitem[{\citenamefont{Prinz et~al.}(1998)}]{Prinz:1998ua}
\bibinfo{author}{\bibfnamefont{A.~A.} \bibnamefont{Prinz}}
  \bibnamefont{et~al.}, \bibinfo{journal}{Phys. Rev. Lett.}
  \textbf{\bibinfo{volume}{81}}, \bibinfo{pages}{1175} (\bibinfo{year}{1998}),
  \eprint{hep-ex/9804008}.

\bibitem[{\citenamefont{Chang et~al.}(2018)\citenamefont{Chang, Essig, and
  McDermott}}]{Chang:2018rso}
\bibinfo{author}{\bibfnamefont{J.~H.} \bibnamefont{Chang}},
  \bibinfo{author}{\bibfnamefont{R.}~\bibnamefont{Essig}}, \bibnamefont{and}
  \bibinfo{author}{\bibfnamefont{S.~D.} \bibnamefont{McDermott}},
  \bibinfo{journal}{JHEP} \textbf{\bibinfo{volume}{09}}, \bibinfo{pages}{051}
  (\bibinfo{year}{2018}), \eprint{1803.00993}.

\bibitem[{\citenamefont{Vogel and Redondo}(2014)}]{Vogel:2013raa}
\bibinfo{author}{\bibfnamefont{H.}~\bibnamefont{Vogel}} \bibnamefont{and}
  \bibinfo{author}{\bibfnamefont{J.}~\bibnamefont{Redondo}},
  \bibinfo{journal}{JCAP} \textbf{\bibinfo{volume}{02}}, \bibinfo{pages}{029}
  (\bibinfo{year}{2014}), \eprint{1311.2600}.

\bibitem[{\citenamefont{Ramani and Woolley}(2019)}]{Ramani:2019jam}
\bibinfo{author}{\bibfnamefont{H.}~\bibnamefont{Ramani}} \bibnamefont{and}
  \bibinfo{author}{\bibfnamefont{G.}~\bibnamefont{Woolley}}
  (\bibinfo{year}{2019}), \eprint{1905.04319}.

\bibitem[{\citenamefont{Bauer et~al.}(2017)\citenamefont{Bauer, Neubert, and
  Thamm}}]{Bauer:2017ris}
\bibinfo{author}{\bibfnamefont{M.}~\bibnamefont{Bauer}},
  \bibinfo{author}{\bibfnamefont{M.}~\bibnamefont{Neubert}}, \bibnamefont{and}
  \bibinfo{author}{\bibfnamefont{A.}~\bibnamefont{Thamm}},
  \bibinfo{journal}{JHEP} \textbf{\bibinfo{volume}{12}}, \bibinfo{pages}{044}
  (\bibinfo{year}{2017}), \eprint{1708.00443}.

\bibitem[{\citenamefont{Carenza et~al.}(2020)\citenamefont{Carenza, Straniero,
  D\"obrich, Giannotti, Lucente, and Mirizzi}}]{Carenza:2020zil}
\bibinfo{author}{\bibfnamefont{P.}~\bibnamefont{Carenza}},
  \bibinfo{author}{\bibfnamefont{O.}~\bibnamefont{Straniero}},
  \bibinfo{author}{\bibfnamefont{B.}~\bibnamefont{D\"obrich}},
  \bibinfo{author}{\bibfnamefont{M.}~\bibnamefont{Giannotti}},
  \bibinfo{author}{\bibfnamefont{G.}~\bibnamefont{Lucente}}, \bibnamefont{and}
  \bibinfo{author}{\bibfnamefont{A.}~\bibnamefont{Mirizzi}},
  \bibinfo{journal}{Phys. Lett. B} \textbf{\bibinfo{volume}{809}},
  \bibinfo{pages}{135709} (\bibinfo{year}{2020}), \eprint{2004.08399}.

\bibitem[{\citenamefont{Depta et~al.}(2020)\citenamefont{Depta, Hufnagel, and
  Schmidt-Hoberg}}]{Depta:2020wmr}
\bibinfo{author}{\bibfnamefont{P.~F.} \bibnamefont{Depta}},
  \bibinfo{author}{\bibfnamefont{M.}~\bibnamefont{Hufnagel}}, \bibnamefont{and}
  \bibinfo{author}{\bibfnamefont{K.}~\bibnamefont{Schmidt-Hoberg}},
  \bibinfo{journal}{JCAP} \textbf{\bibinfo{volume}{05}}, \bibinfo{pages}{009}
  (\bibinfo{year}{2020}), \eprint{2002.08370}.

\end{thebibliography}

\end{document}